\documentclass[times, 10pt,twocolumn]{article}
\usepackage{latex8}
\usepackage{times}

\usepackage{amssymb}
\usepackage{graphicx}

\usepackage{algorithm}
\usepackage{algorithmic}

\newcommand{\QED}{\mbox{\rule[0pt]{1.0ex}{1.0ex}}}
\def\boxend{\hspace*{\fill} $\QED$}
\def\done{\boxend}

\newtheorem{example}{Example}
\newtheorem{theorem}{Theorem}

\newtheorem{property}{Property}

\newtheorem{definition}{Definition}

\newcommand{\nop}[1]{}

\begin{document}

\title{Efficient Skyline Querying with Variable User Preferences on Nominal Attributes}

\author{Raymond Chi-Wing Wong$^1$, Ada Wai-chee Fu$^1$, Jian Pei$^2$,
Yip Sing Ho$^1$, Tai Wong$^1$, Yubao Liu$^3$
\vspace*{0.2cm} \\
\begin{tabular}{c c c}
$^1$ The Chinese University of Hong Kong & $^2$ Simon Fraser University & $^3$ Sun Yat-Sen University \\
\normalsize{cwwong,adafu@cse.cuhk.edu.hk} &
\normalsize{jpei@cs.sfu.ca} & \normalsize{liuyubao@mail.sysu.edu.cn}
\end{tabular}
}

\maketitle
\thispagestyle{empty}

\begin{abstract}
Current skyline evaluation techniques assume a fixed ordering on the
attributes. However, dynamic preferences on nominal attributes are
more realistic in known applications. In order to generate online
response for any such preference issued by a user, we propose two
methods of different characteristics. The first one
is a
semi-materialization method and the second is an adaptive SFS
method. Finally, we conduct experiments to show the
efficiency of our proposed algorithms.
\end{abstract}


\section{Introduction}
\label{sec:related-work}

The skyline operator has emerged as an important summarization
technique for multi-dimensional datasets. Given a set of
$m$-dimensional data points, the \emph{skyline} $S$ is the set of
all points $p$ such that there is no other point $q$ which
\emph{dominates} $p$. $q$ is said to dominate $p$ if $q$ is better
than $p$ in at least one dimension and not worse than $p$ in all
other dimensions. Consider a customer looking for a vacation package
to Cancun using three criteria: price, hotel-class and number of
stops. We know that lower price, higher hotel class and less stops
are more preferable. Thus, if $p$ is in the skyline, then there is
no other package $q$ which has lower price, higher hotel class and
less stops compared with $p$.

Skyline queries have been studied since 1960s in the theory field
where skyline points are known as \emph{Pareto sets} and
\emph{admissible points}~\cite{BS66} or \emph{maximal
vectors}~\cite{BKS+78}. However, earlier algorithms such
as~\cite{BKS+78,BCL90} are inefficient when there are many data
points in a high dimensional space. The problem of skyline queries
was introduced in the database context in~\cite{BKS01}.

Most of the existing studies handle only numeric attributes.
Consider an example as shown in Table~\ref{tab:1nominal} showing a
set of vacation packages with three attributes or
dimensions\footnote{In this paper, we use the terms ``{\em
attribute}'' and ``{\em dimension}'' interchangeably.}, Price,
Hotel-class and Hotel-group. Most existing works consider the first
two attributes which are numeric, where lower price and higher
hotel-class are more preferable. Many efficient methods have been
proposed for so-called full-space skyline queries which return a
set of skyline points in a specific space (a set of dimensions such
as price and hotel-class).
Some representative methods include 
a block nested loop (BNL) algorithm \cite{BKS01}, a sort first
skyline (SFS) algorithm \cite{CGG+03}, a bitmap method \cite{TEO01},
a nearest neighbor (NN) algorithm \cite{KRR02} and a branch and
bound skylines (BBS) method \cite{PTF+03,PTF+05}.
Recently, skyline computation has been extended to consider subspace
skyline queries which return the skylines in
subspaces~\cite{YLL+05,PJE+05,XZ06,PYL06,PFL+07}.

\begin{table*}[t]
\small \center
\begin{tabular}{c c c}
\begin{minipage}[htbp]{6.1cm}
\begin{tabular}{| c | c | c | c |} \hline
  Package & Price & Hotel-class & Hotel-group \\ \hline
  $a$ & 1600 & 4 & T (Tulips) \\ \hline
  $b$ & 2400 & 1 & T (Tulips) \\ \hline
  $c$ & 3000 & 5 & H (Horizon) \\ \hline
  $d$ & 3600 & 4 & H (Horizon) \\ \hline
  $e$ & 2400 & 2 & M (Mozilla) \\ \hline
  $f$ & 3000 & 3 & M (Mozilla) \\ \hline
\end{tabular}
\\
\caption{\label{tab:1nominal} Vacation packages \vspace*{-0.5cm}}
\vspace*{-0.5cm}
    \end{minipage}
& \hspace*{1cm} &
\begin{minipage}[htbp]{6.5cm}
\begin{tabular}{| c | c | c |} \hline
  {Customer} & Preference & Skyline\\ \hline
  Alice & $T \prec M \prec *$ & \{ a, c \} \\ \hline
  Bob & No special preference  & \{ a, c, e, f \} \\ \hline
  Chris & $H \prec M \prec *$ & \{ a, c, e \} \\ \hline
  David & $H \prec M \prec T$ & \{ a, c, e \}  \\ \hline
  Emily & $H \prec T \prec *$ & \{ a, c \}  \\ \hline
  Fred & $M \prec *$  & \{ a, c, e, f \}  \\ \hline
\end{tabular}
\\
\caption{\label{tab:customer} Customer preferences \vspace*{-0.5cm}}
\vspace*{-0.5cm}
    \end{minipage}
\end{tabular}
\end{table*}

Hotel-group as shown in Table~\ref{tab:1nominal} is a categorical
attribute. There can be partial ordering on categorical attributes.
 Some recent
studies~\cite{CET05, CET05b, CDK06, Chomicki06a, Chomicki06b, BGS06,
BTL07, WPF+07a} consider partially-ordered categorical attributes.
In \cite{CET05, CET05b}, each partially-ordered attribute is
transformed into two-integer attributes such that the conventional
skyline algorithms can be applied. \cite{CDK06} studies the cost
estimation of the skyline operator involving the partially ordered
attributes.

Nevertheless, known existing work on categorical attributes assumes
that {\em each attribute has only one order: either a total or a
partial order}. In real life, it is not often that categorical
attributes have a fixed predefined order. For example, different
customers may prefer different realty locations, different car
models, or different airlines. We call such a categorical attribute
which does not come with a predefined order a \emph{nominal
attribute}. It is easy to name important applications with nominal
attributes, such as realties (where type of realty, regions and
style are examples of nominal attributes) and flight booking (where
airline and transition airport are examples of nominal attributes).
In this paper, we consider the scenarios where different users may
have different preferences on nominal attributes. That is, more than
one order need to be considered in nominal attributes.

Furthermore, typically, for a nominal attribute, there may be many
different values, and a user would not specify an order on all the
values, but would only list a few of the most favorite choices.
Table~\ref{tab:customer} shows different customer preferences on
Hotel-group. The preference of Alice is ``$T \prec M \prec
*$" which means that she prefers Tulips to Mozilla and prefers
these two to other hotel groups (i.e., Horizon). We call such
preferences \emph{implicit preferences}. Note that different
preferences yield different skylines. As shown in
Table~\ref{tab:customer}, the skyline is $\{a, c\}$ for Alice's
preference but $\{a, c, e, f\}$ for Fred's preference. The numerous
skylines make the problem highly challenging.

Some latest works \cite{Chomicki06a,Chomicki06b} study the problem
of preference changes, whereupon the query results can be
incrementally refined.
%
   In \cite{BGS06}, a user or a customer can specify some values in nominal attributes
   as an equivalence class to denote the same ``importance" for those values.
%
   \cite{BTL07} is an extension of \cite{BGS06}.
   In \cite{BTL07}, whenever a user finds that there are
   a lot of irrelevant results for a query, s/he can modify the query by
   adding more conditions so that the result
   set is smaller to suit her/his need.
However, these works only focus either on the effects of the query
changes on the result size, or the reuse of skyline results when a
query is refined in a progressive manner, but not on finding
efficient algorithms. Here, we consider that different users may have
different preferences and so the preferences are not undergoing
refinement but they can be different or conflicting from one query
to another. Also, we focus on the issue of efficient query answering.
Nominal attributes are first considered in \cite{WPF+07a} but there
the study is about finding a set of partial orders with respect to
which a given point is in the skyline.

In \cite{PTF+05}, dynamic skyline is considered but it is only for
numeric data, and the ``dynamic function'' considered is based on distance from
a user location. Here, we consider nominal attributes, and the
``dynamic function'' is any mapping between the nominal values and the
\emph{rankings} where each nominal value is assigned with a ranking value.
The BBS method does not work in our case.



Our contributions include the following.
\noindent (1) To the best of our knowledge, this is the first work to
study the problem of
    efficient skyline querying with respect to dynamic implicit preference
    on nominal attributes. 
\noindent (2)  We propose two efficient algorithms of different
flavors, namely
  IPO-Tree Search and Adaptive SFS.
  IPO-Tree is a partial materialization of the skylines for all
  possible implicit preferences. It facilitates the efficient
  computation of the skyline for any implicit preference.
  Adaptive SFS is a little slower but it does not require materialization and has
  the nice properties of being progressive and allows for
  incremental maintenance.
\noindent (3)  We have conducted extensive experiments to show the
    the efficiency of our proposed algorithms.

\section{Problem Definition}\label{sec:prob-defn}

%
%

A skyline analysis involves multiple attributes. A user's preference
on the values in an attribute can be modeled by a partial order on
the attribute. A \textit{partial order} $\preceq$ is a reflexive,
asymmetric and transitive relation. A partial order is also a total
order if, for any two values $u$ and $v$ in the domain, either $u
\preceq v$ or $v \preceq u$. We write $u \prec v$ if $u \preceq v$
and $u \neq v$. A partial order also can be written as $R = \{(u,
v)| u \preceq v\}$. $u \preceq v$ also can be written as $(u, v) \in
R$.
We call this model as the \emph{partial order model}.

By default, we consider points in an $m$-dimensional
space $\mathbb{S}=D_1 \times \cdots
\times D_m$. For each dimension $D_i$, we assume that there is a
partial or total order $R_i$ on the values in $D_i$. For a point
$p$, $p.D_i$ is the projection on dimension $D_i$. If $(p.D_i,
q.D_i) \in R_i$, we also write $p.D_i \preceq q.D_i$.

For points $p$ and $q$, $p$ \textit{dominates} $q$, denoted by $p
\prec q$, if, for any dimension $D_i \in \mathbb{S}$, $p
\preceq_{D_i} q$, and there exists a dimension $D_{i_0} \in
\mathbb{S}$ such that $p \prec_{D_{i_0}} q$. If $p$ dominates $q$,
then $p$ is more preferable than $q$ according to the preference
orders. The \textit{dominance relation} $R$ can be viewed as the
integration of the preference partial orders on all dimensions.
Thus, we can write $R=(R_1, \ldots, R_m)$. It is easy to see that
the dominance relation is a strict partial order.

Given a data set $\mathcal{D}$ containing data points in space
$\mathbb{S}$, a point $p \in \mathcal{D}$ is in the \textit{skyline}
of $\mathcal{D}$ (i.e., a \textit{skyline point} in $\mathcal{D}$)
if $p$ is not dominated by any points in $\mathcal{D}$. Given a
preference $R$, the skyline of $\mathcal{D}$, denoted by $SKY(R)$,
is the set of skyline points in $\mathcal{D}$. \nop{The {\bf problem
of skyline computation} is to compute the skyline of $\mathcal{D}$
for a given data set $\mathcal{D}$ in space $\mathbb{S}$.}

In many applications, there often exist some orders on some of the
dimensions that hold for all users. In our example in
Table~\ref{tab:1nominal}, a lower price and a higher hotel-class are
always more preferred by customers. Even for nominal attributes,
there may exist some
universal partial orders. 
Hence, we
assume that we are given a \textit{template}, which contains a
partial order for every dimension. The partial orders in the
template are applicable to all users. Each user can then express
his/her specific preference by refining the template. The
containment relation of orders captures the refinement.

For partial orders $R$ and $R'$, $R'$ is a \textit{refinement} of
$R$, denoted by $R \subseteq R'$, if for any $(u, v) \in R$, $(u, v)
\in R'$. Moreover, if $R \subseteq R'$ and $R \neq R'$, $R'$ is said
to be \textit{stronger} than $R$.
%
Let $R = \{(T, M)\}$ and $R' = \{(T, M), (H, M)\}$. Then, $R
\subseteq R'$. That is, $R'$ is a refinement of $R$ by adding a
preference $H \prec M$. As $R \neq R'$, $R'$ is stronger than $R$.

\begin{property}\label{ppty:containment} For orders $R=(R_1,
\ldots, R_m)$ and $R'=(R_1'$, $\ldots$, $R_m')$, $R \subseteq R'$ if
and only if $R_i \subseteq R_i'$ for $1 \leq i \leq m$. \boxend
\end{property}

%

\begin{theorem}[Monotonicity]\label{thm:mono} (\cite{WPF+07a}) Given a data set
$\mathcal{D}$ and a template $R$, if $p$ is not in the skyline with
respect to $R$, then $p$ is not in the skyline with respect to any
refinement $R'$ of $R$. \em \done
\end{theorem}

Theorem~\ref{thm:mono} indicates that, when the orders on the
dimensions are strengthened, some skyline points may be
disqualified. However, a non-skyline point never gains the skyline
membership due to a stronger order. This monotonic property greatly
helps in analyzing skylines with respect to various orders.

\begin{definition}[Conflict-free](\cite{WPF+07a}) Let $R$ and $R'$ be two partial
orders. $R$ and $R'$ are \textit{conflict-free} if there exist no
values $u$ and $v$ such that $u \neq v$, $(u,v) \in R$, and $(v,u)
\in R'$.
\end{definition}



Although the model of partial order refinements can model diverse
individual preferences, it does not fit tightly the real world
scenarios. In a skyline query, for a nominal attribute, users
typically would not explicitly order all values, but may specify a
few of their favorite choices and also give them an ordering. For
example, a user may specify that the first choice is $v$, the second
choice is $v'$. The implicit meaning is that $v$ and $v'$ are better
than all the other choices, say $v_1, v_2, ..., v_k$. We can model
this by the partial order model, by including $v
\prec v'$, $v \prec v_1$, $v \prec v_2$, 
..., $v
\prec v_k$ and $v' \prec v_1$,
$v' \prec v_2$, ..., $v' \prec v_k$. 
We denote this preference by ``$v \prec v' \prec *$" where $*$ means
all choices other than $v$ and $v'$ (in this case, $*$ corresponds
to $\{v_1, v_2, ..., v_k\}$). We call this special kind of partial
order an \emph{implicit preference} and assume that it is
represented in such a form.
For example, the implicit preference ``$H \prec M \prec *$"
corresponds to a set of binary orders $\{(H, M), (H, T), (M, T)\}$
in the partial order model.

\begin{definition}[Implicit Preferences]
Let $v_1, v_2, ...v_k$ be all the values in a nominal attribute
$D_i$. An implicit preference $\widetilde{R}_i$ on $D_i$ is given by
$v_1 \prec v_2 \prec ... v_x \prec
*$. It is equivalent to the partial order given by $\{(v_i, v_j) | i <
j \ \wedge \ i \in [1, x] \ \wedge \ j \in [1,k]\}$.
\end{definition}

In the above definition, $\widetilde{R}_i$ is said to be an
\emph{$x$-th order implicit preference}.
Also, the \emph{order} of $\widetilde{R}_i$, denoted by $order(\widetilde{R}_i)$, is defined to be $x$
and the \emph{order} of $\widetilde{R}$ is defined to be $\max_i\{order(\widetilde{R}_i)\}$.
A value $v_j$ is said to be
in $\widetilde{R}_i$ if $v_j \in \{v_1, v_2, ..., v_x\}$. Also,
$v_j$ is said to be the \emph{$j$-th entry} in $\widetilde{R}_i$.
$\mathcal{P}(\widetilde{R}'_i)$ is defined to be $\{(v_i, v_j) | i <
j\mbox{ and }i \in [1, x] \mbox{ and } j \in [1,k]\}$. Let
$\widetilde{R}' = (\widetilde{R}'_1, \widetilde{R}'_2, ...,
\widetilde{R}'_{m})$. $\mathcal{P}(\widetilde{R}')$ is defined to
be $\bigcup_{i=1}^{m}\mathcal{P}(\widetilde{R}'_i)$.

In this paper, we adopt the convention that $\widetilde{R}'$ denotes
an implicit preference and $R'$ denotes a partial order (which may
or may not be an implicit preference). Also we denote $SKY($
$\mathcal{P}(\widetilde{R}')$ $)$ by $SKY(\widetilde{R}')$.

\begin{definition}[Problem]
Given a dataset $\mathcal{D}$ and an implicit preference
$\widetilde{R}'$, find the skyline $SKY(\widetilde{R}')$ in
$\mathcal{D}$. \done
\end{definition}

The problem defined above is our objective in this paper. We also
say that we want to find a set of skyline points with respect to
$\widetilde{R}'$ in $\mathcal{D}$. In many
applications, online response is required. 
The extensive study in \cite{PTF+05} reports that all the existing
algorithms have some serious shortcomings and a new algorithm BBS is
proposed which is much more efficient than previous methods. However,
the data partitioning in BBS is based on fixed orderings on the
dimensions and the same partitioning cannot be used for dynamic or
variable preferences on nominal attributes. Therefore, new mechanisms
need to be explored.

The problem of dynamic implicit preferences have some similar flavor
to subspace skylines since materialization of the possible skylines
seems to be a solution. However, as noted in \cite{PTF+05}, most
applications involve up to five attributes, the dimensionality of a
typical skyline problem is not high, and therefore materialization
of the skylines is quite feasible and has been investigated in
recent works such as \cite{YLL+05,XZ06,PYL06,PFL+07}. For dynamic implicit
preferences, the number of combinations is exponential not only in
the dimensionality but also in the cardinalities of the attributes,
which makes the problem much more challenging.


\begin{table}[t]
\small  \begin{center}
\begin{minipage}{85mm}
\begin{tabular}{|c|c|c|c|c|} \hline
  \parbox[t]{1cm}{Package } & Price & Hotel-class & Hotel group & Airline\\ \hline
  $a$ & 1600 & 4 & T (Tulips) & G (Gonna) \\ \hline
  $b$ & 2400 & 1 & T (Tulips) & G (Gonna) \\ \hline
  $c$ & 3000 & 5 & H (Horizon) & G (Gonna) \\ \hline
  $d$ & 3600 & 4 & H (Horizon) & R (Redish) \\ \hline
  $e$ & 2400 & 2 & M (Mozilla) & R (Redish) \\ \hline
  $f$ & 3000 & 3 & M (Mozilla) & W (Wings) \\ \hline
\end{tabular}
\\
\caption{\label{tab:2nominal} A table with two nominal attributes.
\vspace*{-0.5cm}} \vspace*{-0.5cm}
\end{minipage}
\end{center}
\end{table}

\section{Partial Materialization: IPO-Tree Search}
\label{sec:implicit}

In order to support online response, a naive approach is to
materialize the skylines for all possible preferences. However, as
noted in the above, this approach is very costly in storage and
preprocessing. Our study in \cite{WPF+07b} shows that, even with an
index and with compression by removing redundancies in shared
skylines, the cost is still prohibitive.


Our idea is therefore to materialize some useful partial results so
that these partial results can be combined efficiently to form the
query results. In particular, we propose to materialize the results
with respect to the first-order implicit preference on each nominal
attribute only. Since results for the second or higher order
preferences are not stored, the number of combinations is
significantly reduced. In the following, we describe an important
property called the \emph{merging property} which allows us to
derive results of all possible implicit preferences of \emph{any}
order by simple operations on top of the \emph{first}-order
information maintained.

\begin{theorem}[Merging Property]
Let two implicit preferences $\widetilde{R}'$ and $\widetilde{R}''$
differ only at the $i$-th dimension, i.e.,
$\widetilde{R}_j' = \widetilde{R}_j''$ for all $j \neq i$.
Furthermore, $\widetilde{R}_i' = $``$v_1 \prec ... \prec v_{x-1}
\prec *$" and $\widetilde{R}_i'' = $``$v_{x} \prec
*$". Let
$PSKY(\widetilde{R}')$ be the set of points in $SKY(\widetilde{R}')$
with $D_i$ values in $\{v_1, ... v_{x-1}\}$. Let $\widetilde{R}_i'''
= $``$v_1 \prec ... \prec v_{x-1} \prec v_{x} \prec
*$". The skyline with respect to $\widetilde{R}'''$ is $(SKY(\widetilde{R}')
\cap SKY(\widetilde{R}'')) \cup PSKY(\widetilde{R}')$. \label{thm:merging2}

\bigskip

\em \noindent\textbf{Proof:} A proof is given in the Appendix. \done
\end{theorem}

\begin{figure}[t]
         \centerline{\includegraphics[width=6cm]{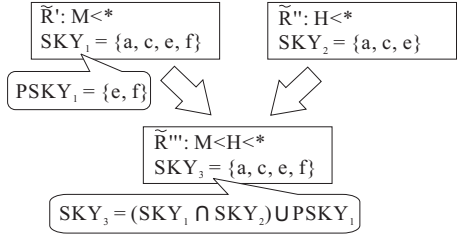}}
         \caption{Illustration of the merging property \vspace*{-0.3cm}}
         \vspace*{-0.3cm}
         \label{fig:mergeSkyline}
\end{figure}

For example, in Figure~\ref{fig:mergeSkyline}, let $\widetilde{R}'$
be ``$M \prec
*$" and $\widetilde{R}''$ be ``$H \prec *$". From
Table~\ref{tab:1nominal}, the skyline with respect to
$\widetilde{R}'$ is $SKY_1 = \{a, c, e, f\}$ and the skyline with
respect to $\widetilde{R}''$ is $SKY_2 = \{a, c, e\}$. $PSKY_1 =
\{e,f\}$ is the set of skyline points with values in $\{M\}$.
 Let $\widetilde{R}'''$ be ``$M \prec H \prec
*$". By Theorem~\ref{thm:merging2}, the skyline $SKY_3$ with
respect to $\widetilde{R}'''$ is obtained as follows.
$SKY_3 = (SKY_1 \cap SKY_2) \cup PSKY_1$
= $(\{a, c, e, f\} \cap \{a, c, e\}) \cup \{e, f\} = \{a, c, e\}
\cup \{e, f\} = \{a, c, e, f\}$. The derivation can be explained as
follows. $\mathcal{P}(\widetilde{R}')$ and
$\mathcal{P}(\widetilde{R}'')$ are not conflict-free
because their union contains both $(M, H)$ and $(H, M)$. 
Or, the only difference between $\mathcal{P}(\widetilde{R}') \cup
\mathcal{P}(\widetilde{R}'')$ and $\mathcal{P}(\widetilde{R}''')$ is
that $\mathcal{P}(\widetilde{R}') \cup \mathcal{P}(\widetilde{R}'')$
contains one more binary entry, namely $(H, M)$, which may
disqualify some data points (in this example, it disqualifies $f$).
In order to remove the disqualifying effect, we augment the
intersection $SKY_1 \cap SKY_2$ by a union
with $PSKY_1$ where $PSKY_1$ contains 
 the points disqualified by $(H, M)$ in $SKY_1$.

From Theorem \ref{thm:merging2}, we can derive a powerful tool for
the computation of the skyline with respect to any implicit
preference of any order by building increasingly higher order
refinement ($\widetilde{R}'''$ in the theorem) skyline from lower
order ($\widetilde{R}'$ and $\widetilde{R}''$) ones, starting with
the first-order.
In the following two subsections, we introduce the IPO-tree for
storing the first-order preference skylines and the query evaluation
based on the IPO-tree.

\subsection{Tree Construction}

\begin{figure*}[tb] 
\begin{tabular}{c c c}
    \begin{minipage}[t]{10.5cm}
        \centerline{\includegraphics[width=10.5cm]{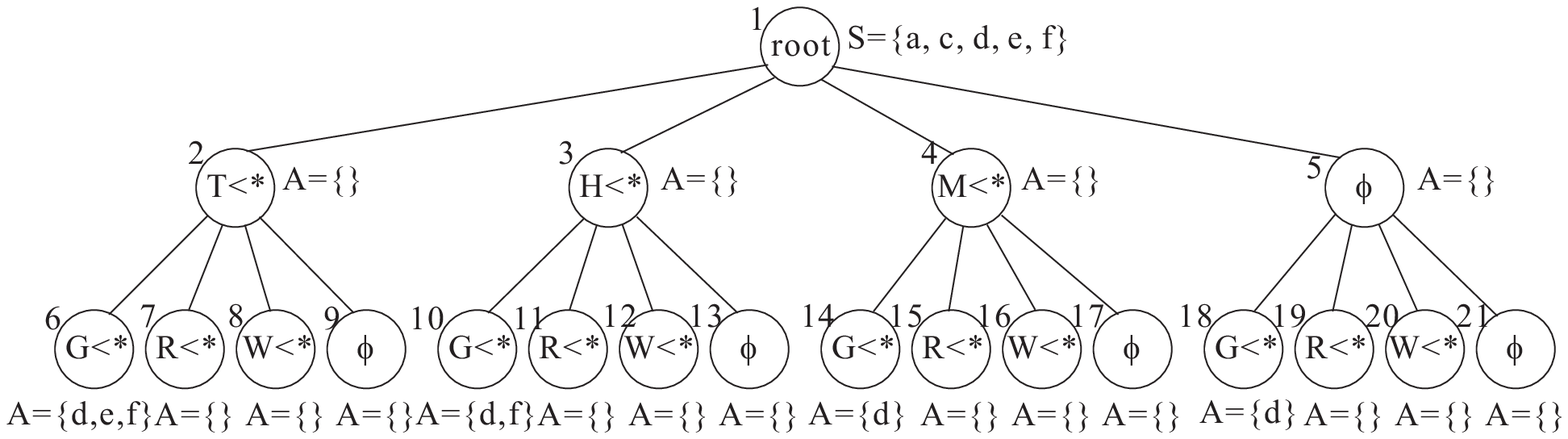}}
       \caption{Illustration of an implicit preference order tree (IPO-tree) \vspace*{-0.3cm}}
       \vspace*{-0.3cm}
      \label{fig:tree}
    \end{minipage}
    &
    \begin{minipage}[t]{6.5cm}
       \centerline{\includegraphics[width=6.5cm]{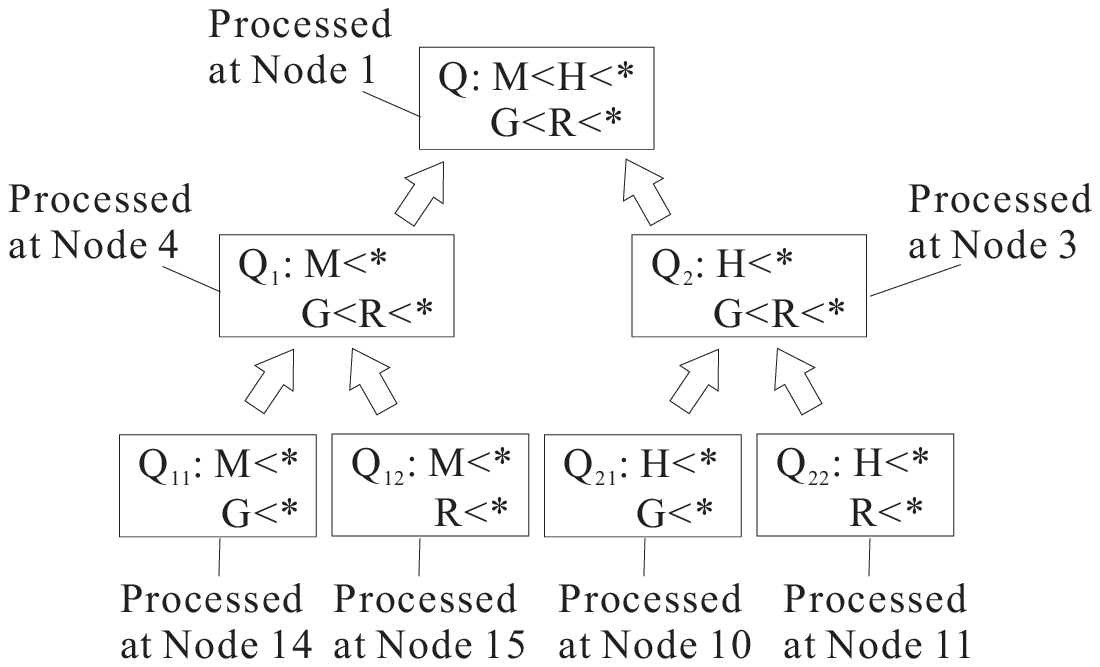} }
      \caption{Query evaluation with an IPO-tree \vspace*{-0.3cm}}
      \vspace*{-0.3cm}
      \label{fig:mergeSkylineComplicatedWithTree}
    \end{minipage}
\end{tabular}
\end{figure*}

An \emph{IPO-tree} (\emph{implicit preference order tree}) stores
results for combinations of first-order preferences. In this tree,
each node is labeled with a first-order implicit preference, namely
``$v \prec
*$", where $v \in D_i$ and $D_i$ is a nominal dimension. 
The tree is of depth $m' + 1$, where $m'$ is the number of nominal
attributes. The root node stores the skyline $SKY(R)$ with respect
to template $R$ in $\mathcal{D}$. The second level contains all nodes
corresponding to first-order implicit preferences on nominal
attribute $D_1$. In general, the children of an $i$-th level node
correspond to all the first-order implicit preferences on nominal
attribute $D_{i}$. A special child node is labeled $\phi$
corresponding to no preference.
Each non-root node has a label associated with a first-order
implicit preference on a single nominal attribute, and maintains
\emph{results} that corresponds to the labels along the path to the root
node. Figure~\ref{fig:tree} shows an IPO-tree from the data in
Table~\ref{tab:2nominal}, where the template $R$ is set to
$\emptyset$.
Node 6 corresponds to implicit preferences ``$T \prec *, G \prec
*$".

Furthermore, a root node is associated with a set $S = SKY(R)$.
But, each non-root node is associated with a set $\cal A$ of
points where $S - {\cal A}$ is the skyline for the corresponding
implicit preference. Therefore, $\cal A$ contains the points in
$SKY(R)$ that are \emph{disqualified} from the skyline at the
node because of the preference refinement. For example, since, in the
IPO-tree shown in Figure~\ref{fig:tree}, Node 6 corresponds to an implicit
preference ``$T \prec *, G \prec *$", which disqualifies points
$d,e,f$ in $S$ as skyline points, $\cal A$ of node 6 is equal to
$\{d, e, f\}$. The purpose of $\cal A$ is to allow us to find the
skyline for the node given the skylines of the ancestors. It is also
possible to store the exact skyline at each node instead.

\smallskip\noindent\textbf{Implementation:} In order to find the set $\cal A$ for each non-root node $N$, one can apply a
skyline algorithm (e.g., adaptive SFS in Section~\ref{sec:sfs}).
However, in our implementation, we make use of the \emph{minimal
disqualifying conditions} introduced in \cite{WPF+07a}. For a skyline
point $p$ and a template order $R$, a partial order $R'$ is called a
\textit{minimal disqualifying condition} (or MDC for short) if (1)
$R' \cap R = \emptyset$, (2) $R'$ and $R$ are conflict-free, (3) $p$
is not a skyline point with respect to $R \cup R'$, and (4) there
exists no $R''$ such that $R'' \subset R'$ and $p$ is not a skyline
point with respect to $R \cup R''$. The set of minimal disqualifying
conditions
for $p$ is denoted by $MDC(p)$. 
The first step here is to find all MDCs of each skyline point in
$SKY(R)$. One of the algorithms in \cite{WPF+07a} can be used for
this step. Then, given the implicit preference $\widetilde{R}'$
corresponding to a node $N$, we check each point in $SKY(R)$, if any
of the MDCs is a subset of $\cal P (\widetilde{R}')$, then the point
is disqualified and is inserted into $\cal A$.



\smallskip\noindent\textbf{Tree Size:} Let $m'$ be the number of nominal attributes
and $c$ be the maximum
cardinality of a nominal attribute. The height of the IPO-tree is
$m' + 1$. The size of the tree in number of nodes is given by
$O(c^{m'})$. 
As
claimed in \cite{KRR02} and quoted in \cite{PTF+05}, most
applications involve up to five attributes, and hence $m'$ is very
small. Note that the IPO-tree size is significantly smaller than the
number of possible implicit preferences which is given by
$O((c\cdot c!)^{m'})$.

The tree size can be further controlled if we know the query
pattern (e.g., from a history of user queries). Typically, there are
popular and unpopular values. For values which are seldom or never
chosen in implicit preferences, the corresponding tree nodes in the
IPO-tree are not needed. It is possible to restrict the IPO-tree to
say the 10 most popular values for each nominal attribute. If a
query containing unpopular values arrives, the adaptive SFS
algorithm in Section \ref{sec:sfs} can be used instead.

\subsection{Query Evaluation}
\label{subsec:implicitQuery}

IPO-tree has a nice structure with a well-controlled tree size and
can efficiently facilitate implicit preference querying based on
the merging property (Theorem~\ref{thm:merging2}). 
%
Algorithm~\ref{alg:preferenceQuery} shows the evaluation of a query
with an implicit preference $\widetilde{R}'$. 

\begin{algorithm}[h]
\small
\caption{\textbf{query}($d$, $\widetilde{R}'$, $N$, $S$) }
\label{alg:preferenceQuery}
\begin{algorithmic} [1]
\REQUIRE dimension $d$, implicit preference $\widetilde{R}'$, tree node $N$,
set of potential skyline points $S$
\\
Local variable: $\mathcal{Q}$ - a queue containing sets of points
\STATE $X \leftarrow S$
\IF{$d \neq m'$}
  \IF{$R'_{d}$ contains no preferences}
     \STATE $N_c \leftarrow$ the child node of $N$ labeled $\phi$
     \STATE $X \leftarrow $\textbf{query}($d+1$, $\widetilde{R}'$, $N_c$, $S$)
  \ELSE
     \STATE $\mathcal{Q} \leftarrow \emptyset$
     \FOR{$i := 1$ to $order(\widetilde{R}'_{d})$}
        \STATE $v \leftarrow $the $i$-th entry in $\widetilde{R}'_{d}$
        \STATE $N_c \leftarrow$ child node of $N$ labeled with ``$v \prec *$"
        \STATE $\mathcal{A} \leftarrow$ the disqualifying set of $N_c$
        \STATE $Y \leftarrow $\textbf{query}($d+1$, $\widetilde{R}'$, $N_c$, $S - \mathcal{A}$)
        \STATE enqueue $Y$ to $\mathcal{Q}$
     \ENDFOR
     \STATE $X \leftarrow $ \textbf{merge}($d+1$, $\mathcal{Q}$, $\widetilde{R}'$) (See Algorithm~\ref{alg:merge})
  \ENDIF
\ENDIF
\STATE \textbf{return} $X$
\end{algorithmic}
\end{algorithm}
\begin{algorithm}[h]
\small
\caption{\textbf{merge}($d$, $\mathcal{Q}$, $\widetilde{R}'$)}
\label{alg:merge}
\begin{algorithmic} [1]
\REQUIRE dimension $d$, $\mathcal{Q}$ storing sets of points, preference
$\widetilde{R}'$
\STATE dequeue $\mathcal{Q}$ and obtain the dequeued element
$Y$ \STATE $X \leftarrow Y$
\FOR{$i := 2$ to $order(\widetilde{R}'_{d})$}
  \STATE dequeue $\mathcal{Q}$ and obtain the dequeued element $Y$
  \STATE let $\mathcal{R}$ be the set of the first to the ($i-1$)-th entries in $\widetilde{R}'_{d}$
  \STATE $Z \leftarrow$ a set of points $p$ in $X$ with $p.D_d \in \mathcal{R}$
  \STATE $X \leftarrow (X \cap Y)\cup Z$
\ENDFOR
\end{algorithmic}
\end{algorithm}

\begin{example}[Query Evaluation] \em
We use the IPO-tree in Figure~\ref{fig:tree} for the illustration of
the detailed steps in implicit preference query evaluation. Let us
consider four different queries for illustration, namely $Q_A:$ ``$M
\prec
*$", $Q_B:$ ``$M \prec *, G \prec *$", 
$Q_C:$ ``$M \prec H \prec *, G \prec *$" and $Q_D:$ ``$M \prec H
\prec *, G \prec R \prec *$".

Consider $Q_A$. We first visit Node 1 and $X$ is set to be $S$ of
Node 1 (i.e., $\{a, c, d, e, f\}$). Node 4 is then visited where
$\cal A$ is $\emptyset$, $X$ is still $\{a, c, d, e, f\}$, which is
the skyline for $Q_A$.

Consider $Q_B$. After visiting Node 1, $X =
\{a, c, d, e, f\}$. Next, Node 4 and Node 14 are visited. 
The skyline is $X = \{a, c, d, e, f\} - \{d\} = \{a, c, e, f\}$.


Consider $Q_C$. We split the query into subqueries ``$M \prec *, G
\prec
*$" and ``$H \prec *, G \prec *$", with respective skylines of
 $\{a, c, e, f\}$ and $\{a , c, e\}$. The subset $PSKY_1$ of $SKY_1$ with
Hotel-group value $M$ is $\{e, f\}$. By Theorem~\ref{thm:merging2},
the resulting skyline is $(\{a, c, e, f\} \cap \{a, c, e\}) \cup
\{e, f\} = \{a, c, e, f\}$.

Consider $Q_D$. As illustrated in
Figure~\ref{fig:mergeSkylineComplicatedWithTree}, we follow the
breakdown and obtain the skyline with respect to $Q_D$ equal to
$\{a, c, e, f\}$. \done
\end{example}

\begin{theorem}
With Algorithm~\ref{alg:preferenceQuery}, \textbf{query}(1,
$\widetilde{R}'$, $Root$, SKY(R)) returns $SKY(\widetilde{R}')$,
given a template $R$ for a dataset $\mathcal{D}$ and the
corresponding IPO-tree with a root node of $Root$. \done
\end{theorem}

The number of leaf nodes in a query evaluation tree diagram as the
one shown in Figure \ref{fig:mergeSkylineComplicatedWithTree} gives
a bound on the number of set operations. The number of set
operations required for an $x$-th order implicit preference is
$O(x^{m'})$. Since $x$ and $m'$ are very small, this number is also
small.




\smallskip\noindent\textbf{Implementation:} We have implemented the algorithm by accumulating the set of
disqualified points. By Theorem~\ref{thm:merging2}, if $A(\widetilde{R}')$ and
$A(\widetilde{R}'')$ are the sets of disqualified points for $\widetilde{R}'$
and $\widetilde{R}''$, respectively, let $\mathcal{B}$ be the set of
points in $A(\widetilde{R}'')$ with $D_i$ values in $\{v_1, ..,
v_{x-1}\}$, the accumulated set of disqualified points for $\widetilde{R}'''$
is given by
$A(\widetilde{R}') \cup (A(\widetilde{R}'') - \mathcal{B})$.

Another efficient implementation is to store the skyline for each
node in the IPO-tree by means of a bitmap (replacing $\cal A$) and
to create an inverted list for each nominal attribute for an easy lookup
to determine a bitmap for $PSKY(\widetilde{R}')$ (see Theorem
\ref{thm:merging2}). Efficient bitwise operations can then be used
for the set operations.

\section{Progressive Algorithm: Adaptive SFS}
\label{sec:sfs}

The IPO-tree method requires much preprocessing cost and storage. It
is also more appropriate for more static datasets since changes in
the datasets require rebuilding the entries in the tree. It is of
interest to find an efficient algorithm which does not involve major
overheads, and in addition allows incremental maintenance to
accommodate dynamic updating of the datasets. Here, we propose such
a method for real-time querying which is based on the Sort-First
Skyline Algorithm (SFS) \cite{CGG+03}. The algorithm is called
Adaptive SFS and is efficient since it does not require a complete
resorting of the data for each different user preference. It also
allows skyline points to be returned in a progressive manner.

\subsection{Overview of SFS}

First, we will briefly describe the method of Sort-First Skyline
(SFS), which is for totally-ordered numerical attributes. With SFS,
the data points are sorted according to their scores obtained by a
preference function $f$, which can be the sum of all the numeric
values in different dimensions of a data point. That is, the score
of a point $p$ is $
  f(p) = \sum_{i=1}^{m} p.D_i
$. The criterion for the function is that if $p \prec q$, then $f(p)
< f(q)$. The data points are then examined in ascending order of
their scores. A \emph{skyline list} $L$ is initially empty. If a
point is not dominated by any point in $L$, then it is inserted into $L$.
The sorting takes $O(N \log N)$ time while the scanning of the
sorted list to generate the skyline points takes $O(N \cdot n)$
time, where $N$ is the number of data points in the data set and $n$
is the size of the skyline.

\subsection{Adaptive SFS for Implicit Preferences}

Next, we develop an adaptive SFS method for query processing in the
data set with implicit preferences on nominal attributes, given the
skyline set $SKY(\widetilde{R})$ for a template order
$\widetilde{R}$ which is implicit. Let $\widetilde{R}'$ be an
implicit refinement over $\widetilde{R}$. From
Theorem~\ref{thm:mono}, any skyline point $p$ for $\widetilde{R}'$
will also be a skyline point for $\widetilde{R}$. Hence, in order to
look for the skyline for $\widetilde{R}'$, we only need to search
$SKY(\widetilde{R})$.


Our idea is the following. We adopt the basic presorting step on
$SKY(\widetilde{R})$ resulting in a sorted list $L(\widetilde{R})$. When a query
with a refinement $\widetilde{R}'$ arrives, we first try to re-sort
the list $L(\widetilde{R})$ and obtain a new sorted list
$L(\widetilde{R}')$. The skyline generation step is then applied on
$L(\widetilde{R}')$. The key to the efficiency is that the resorting
step complexity is $O(l \log n)$, where $l$ is the number of data
points affected by the refinement $\widetilde{R}'$ and is typically
much smaller than $n$. Next, we give more detailed description of
the algorithm.

Each value $v$ in a dimension $D_i$ is associated with a rank
denoted by $r(v)$. In a totally-ordered attribute $D_i$, we define
$r(v) = v$ for each $v$ in $D_i$. Without loss of generality, we
assume that a smaller value in a dimension $D_i$ is more preferable
than a larger value in the same dimension. For a nominal attribute
$D_i$, we assign $r(v)$ as follows. Let $c_i$ be the cardinality of
nominal dimension $D_i$. By default, for each value $v$ for
dimension $D_i$, $r(v) = c_i$. For example, if there are 10
different values in dimension $D_i$, then by default $r(v) = 10$ for
each $v$ in $D_i$. Given an implicit partial order
$\widetilde{R}'_i$, we can determine a ranking for the values that
appear in $\widetilde{R}'_i$ so that $r(v) < r(v')$ if and only if
$v \prec v'$ can be derived from $\widetilde{R}'_i$. If
$\widetilde{R}'_i$ is ``$v_1 \prec v_2 \prec  ... \prec v_x \prec
* $", then we set $r(v_1) =1$, $r(v_2) = 2$, ..., $r(v_x)=x$.
We define $ f(p) = \sum_{i=1}^{m} r(p.D_i) $.

Let $l$ be the
number of data points that contain some values in $\widetilde{R}'$. The
processing time of the sorting list is $O(l \log n)$.
Algorithms~\ref{preprocess} and \ref{qprocess} show the steps for preprocessing
the data points and
query processing, respectively.


\begin{algorithm}[tb]
\small \caption{\textbf{Preprocessing}} \label{preprocess}
\begin{algorithmic}[1]
\STATE Compute the skyline set $SKY(\widetilde{R})$ for the given template $\widetilde{R}$
\STATE
Determine the ranking $r$ based on $SKY(\widetilde{R})$ and $f$ 
\STATE
 Apply the presorting step of SFS based on $r$ 
 on
 $SKY(\widetilde{R})$
\end{algorithmic}
\end{algorithm}


\begin{algorithm}[tb]
\small \caption{\textbf{Query Processing}} \label{qprocess}
\begin{algorithmic}[1]
\REQUIRE skyline query, with implicit preference $\widetilde{R}'$
 \STATE Determine the
ranking for the values in $\widetilde{R}'$ \STATE Find the data
points in $SKY(\widetilde{R})$ that contain values in
$\widetilde{R}'$. Alter the rankings for such data points if
necessary \STATE Delete the points with altered rankings from the
sorted list \STATE Re-insert the points just deleted 
using the new ranking \STATE Apply the skyline extraction step
of SFS on the resulting sorted list
\end{algorithmic}
\end{algorithm}

In Step 2 in Algorithm~\ref{qprocess}, in order to find data points
in $SKY(\widetilde{R})$ that contain values in $\widetilde{R}'$, one
possible way is to have an index for each nominal dimension. The
index can be a simple sorted list or a more sophisticated tree
index. An index lookup can quickly return the points that contain a
particular value in $\widetilde{R}'$. Such data points are collected
in a set. Then, for each point $p$ in the set, the value of $f(p)$
based on $\widetilde{R}$ allows us to quickly locate the point in
the sorted list. The point is deleted from the list and re-inserted
with a new value for $f(p)$ based on the refinement
$\widetilde{R}'$.

For the last step of the query processing, there is no need to
follow the SFS from scratch. Instead, we reinsert the points in the
ascending order of the new $f(p)$ values. When a point $a$ is
re-inserted, we need only check if it may be dominated by the
$\widetilde{R}'$ skyline points sorted before it. If so, $a$ is not
added; otherwise, 
we then check if it may dominate any $SKY(\widetilde{R})$ skyline
point that are sorted after it. The points that it dominates will be
removed. Let $c = |SKY(\widetilde{R}')|$, $n =
|SKY(\widetilde{R})|$, and $l$ be the number of points in
$SKY(\widetilde{R})$ containing values in $\widetilde{R}'$. The time
complexity of this step will become $O(l \log l + c \cdot l +
\min(c, l) \cdot n)$. Since the resorting step takes $O(l \log n)$
time, the total time is $O(l \log n + \min(c,l) \cdot n)$.

\subsection{Properties of Adaptive SFS}

The presorting ensures that a point $p$ dominating another point $q$
must be visited before $q$. This leads to a \emph{progressive}
behavior, meaning that any point inserted into the skyline list $L$
must be in the skyline set, and it can be reported immediately. The
presorting also enhances the pruning since it is more likely that
candidate points with lower
scores dominate more other points. 
Another desirable property of adaptive SFS is that it allows
\emph{incremental maintenance}. Assume that the algorithm which
finds $SKY(\widetilde{R})$ is incremental. After data is updated,
the set $SKY(\widetilde{R})$ is modified. The sorted list in the
method is altered by simple insertions or deletions. The time
complexity is $O(\log n)$ for each such update.

\section{Empirical Study}

We have conducted extensive experiments on a Pentium IV 3.2GHz PC
with 2GB memory, on a Linux platform. The algorithms were
implemented in C/C++. In our experiments, we adopted the data set
generator released by the authors of \cite{WPF+07a}, which contains
both numeric attributes and nominal attributes, where
the nominal attributes are generated according to a Zipfian
distribution. 
%
The default values of the experimental parameters are shown in
Table~\ref{tab:default}. In the experiment, if the order of the
implicit preference $\widetilde{R}'$ is set to $x$, it means that
the order of $\widetilde{R}'_i$ for each nominal attribute $D_i$ is
$x$.
Note that the total number of dimensions is equal to the number of
numeric dimensions plus the number of nominal dimensions.
%
   By default, we adopted a template where the most frequent value in a
   nominal dimension has a higher preference than all other
   values. This corresponds to a more difficult setting as the skyline tends to be bigger.
   In the following, we use the default settings unless
   specified otherwise.

\begin{table}
\small \begin{center}
\begin{tabular}{| l | c |}\hline
  Parameter & Default value \\ \hline
  No. of tuples & 500K \\ \hline
  No. of numeric dimensions & 3 \\ \hline
  No. of nominal dimensions & 2 \\ \hline
  No. of values in a nominal dimension & 20 \\ \hline
  Zipfian parameter $\theta$ & 1 \\ \hline
  order of implicit preference & 3 \\ \hline
\end{tabular}
\end{center}
\caption{Default values} \label{tab:default}
\end{table}

We denote our proposed partial materialization
methods (IPO Tree Search)
by 
\emph{IPO Tree} and \emph{IPO Tree-10} where \emph{IPO Tree} is
constructed based on all possible nominal values and \emph{IPO
Tree-10} is constructed based on only the 10 most frequent values
for each nominal attribute. We denote the Adaptive SFS algorithm by
\emph{SFS-A}. We also compare our proposed methods with a baseline
algorithm called \emph{SFS-D}, which is the original SFS algorithm
\cite{CGG+03} returning $SKY(\widetilde{R}')$ with respect to
implicit preference $\widetilde{R}'$ for dataset $\mathcal{D}$.

We evaluate the performance of the algorithms in terms of (1)
pre-processing time, (2) the query time of an implicit preference
and (3) memory requirement. We also report (4) the proportion of the
skyline points with respect to the template $\widetilde{R}$ (i.e.,
$|SKY(R)|/|\mathcal{D}|$), (5) the proportion of skyline points
affected in $SKY(\widetilde{R})$ with respect to $\widetilde{R}'$
(i.e., $|AFFECT(R)|/|SKY(R)|$), where $AFFECT(R)$ is the set of
skyline points in $SKY(\widetilde{R})$ with values in
$\widetilde{R}'$, and (6) the proportion of skyline points with
respect to $\widetilde{R}'$ in $SKY(\widetilde{R})$ (i.e.,
$|SKY(R')|/|SKY(R)|$). For pre-processing, both \emph{IPO Tree} and
\emph{IPO Tree-10} compute $SKY(\widetilde{R})$ and build the
correspondence IPO trees, and \emph{SFS-A} compute
$SKY(\widetilde{R})$ and pre-sort the data according to the
preference function $f$. Note that \emph{SFS-D} does not require any
preprocessing. The storage of \emph{IPO Tree} or \emph{IPO Tree-10}
corresponds to the IPO tree stored. \emph{SFS-A} stores the sorted
data in $SKY(\widetilde{R})$, and \emph{SFS-D} does not use extra
storage but reads the data directly from the dataset.

For measurements (1) and (3), each experiment was conducted 100
times and the average of the results was reported. For measurements
(2), (4), (5) and (6), in each experiment, we randomly generated 100 implicit
preferences, and the average query time is reported. We will study
the effects of varying (1) database size, (2) dimensionality, (3)
cardinality of nominal attribute and (4) order of implicit preference.

\subsection{Synthetic Data Set}

Three types of data sets are generated as described in \cite{BKS01}:
(1)
\emph{independent data sets}, 
(2)
\emph{correlated data sets} and 
(3)
\emph{anti-correlated data sets}. 
%
The detailed description of these data sets can be found in
\cite{BKS01}. For interest of space, we only show the experimental
results for the anti-correlated data sets. The results for the
independent data sets and the correlated data sets are similar in
the trend but their execution times are much shorter.

\begin{figure*}[tbp] 
\begin{tabular}{c c c c}
    \begin{minipage}[htbp]{4.0cm}
        \includegraphics[width=4.0cm,height=3.0cm]{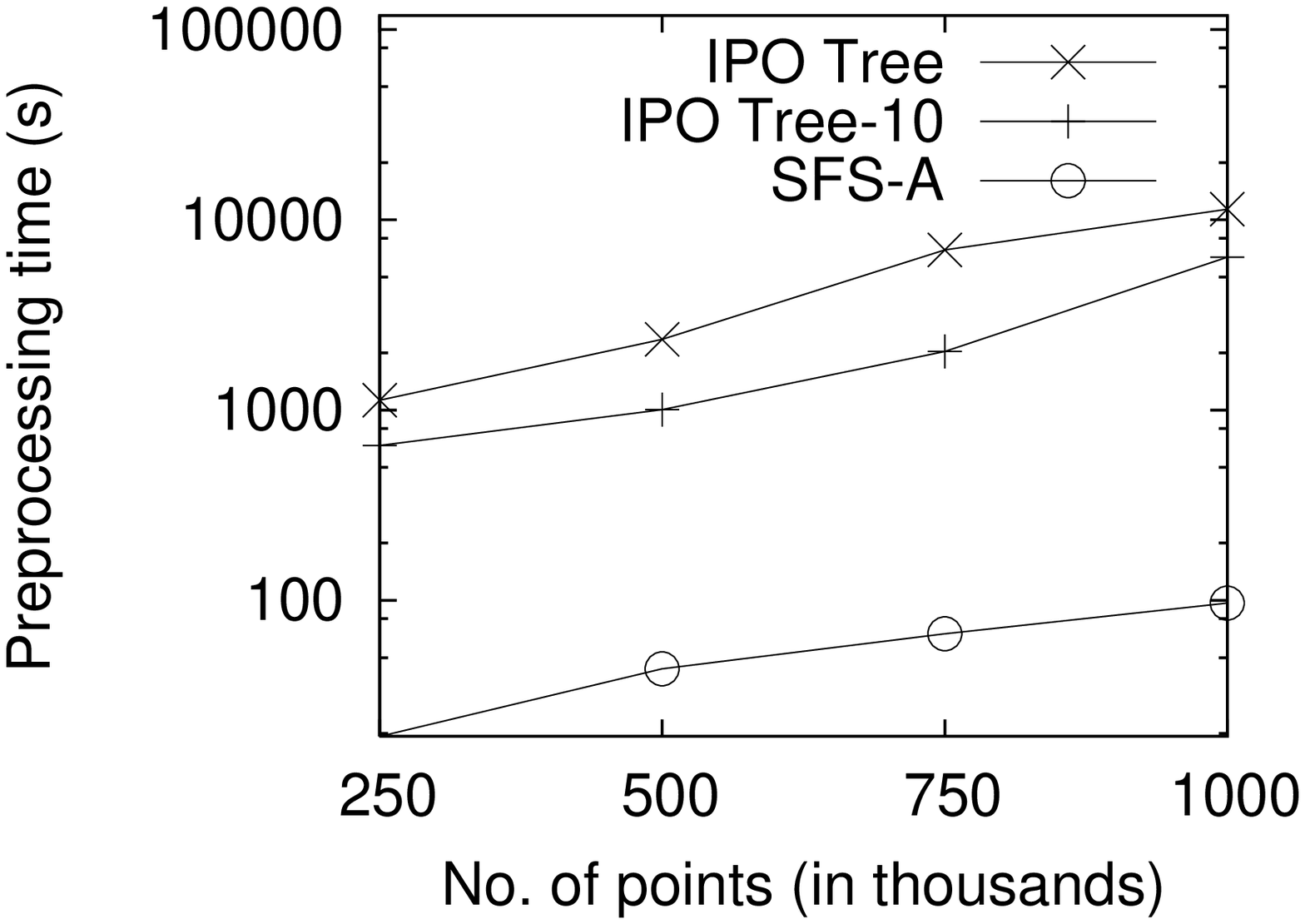}
    \end{minipage}
&
    \begin{minipage}[htbp]{4.0cm}
        \includegraphics[width=4.0cm,height=3.0cm]{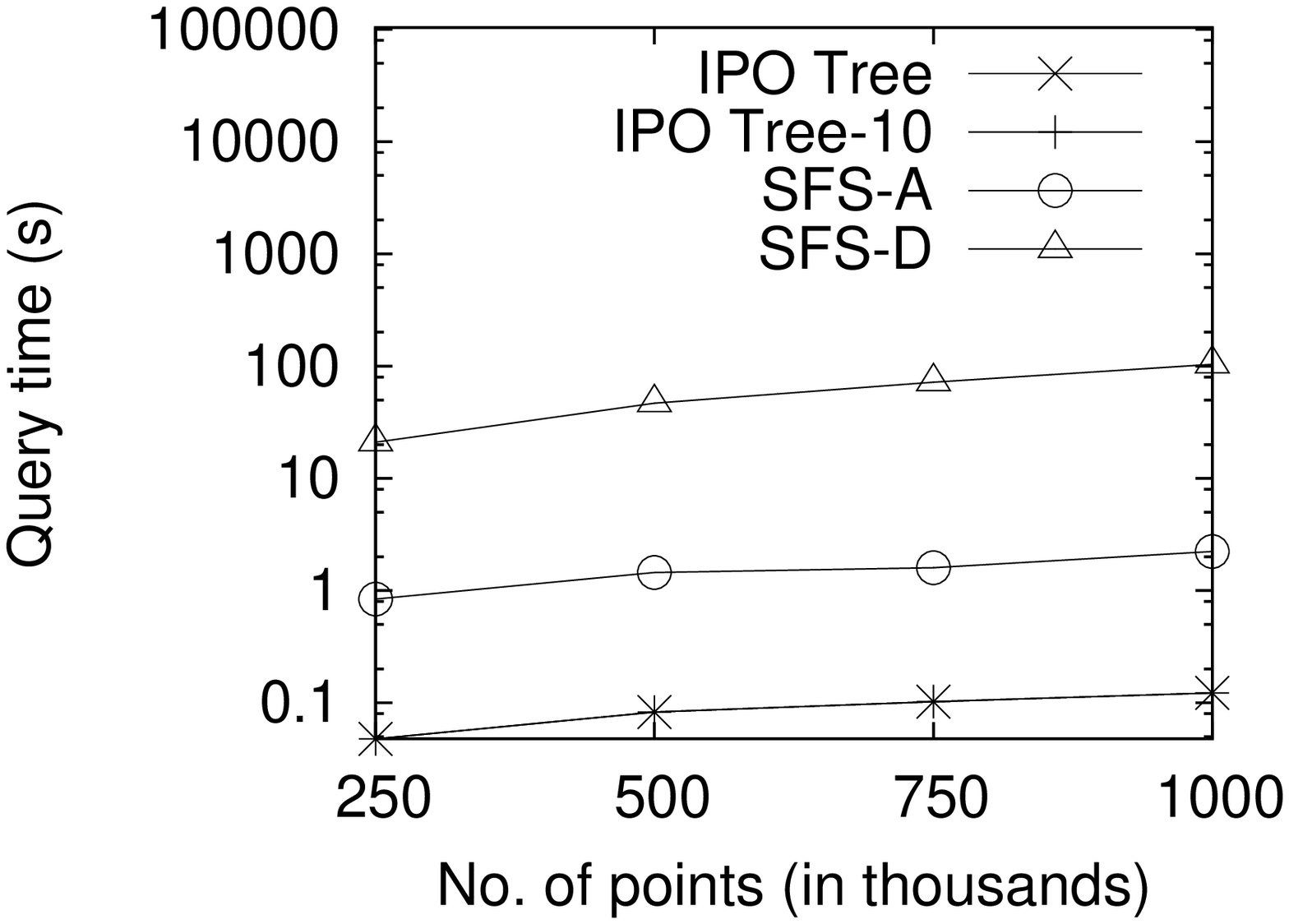}
    \end{minipage}
&
    \begin{minipage}[htbp]{4.0cm}
        \includegraphics[width=4.0cm,height=3.0cm]{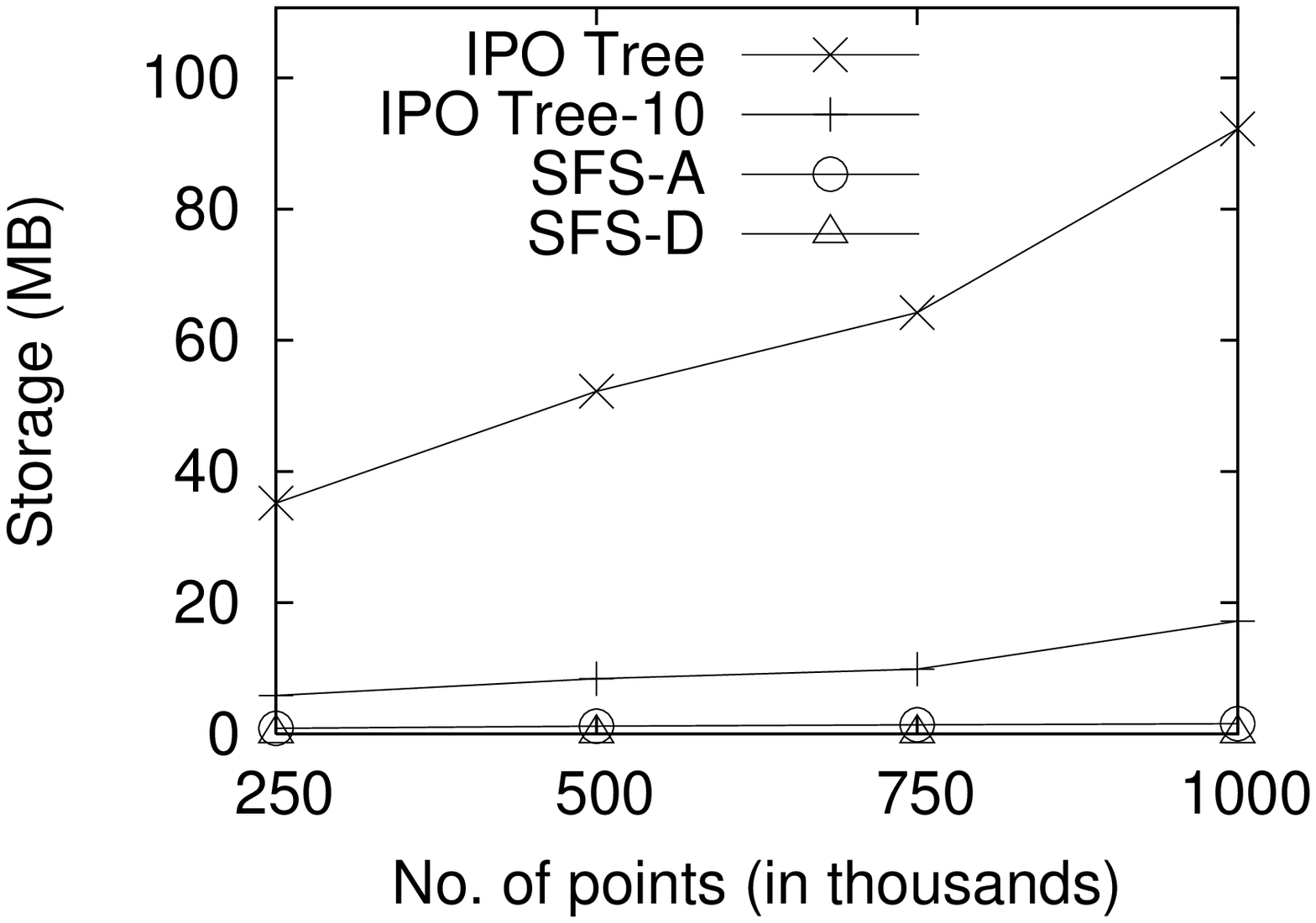}
    \end{minipage}
&
    \begin{minipage}[htbp]{4.0cm}
        \includegraphics[width=4.0cm,height=3.0cm]{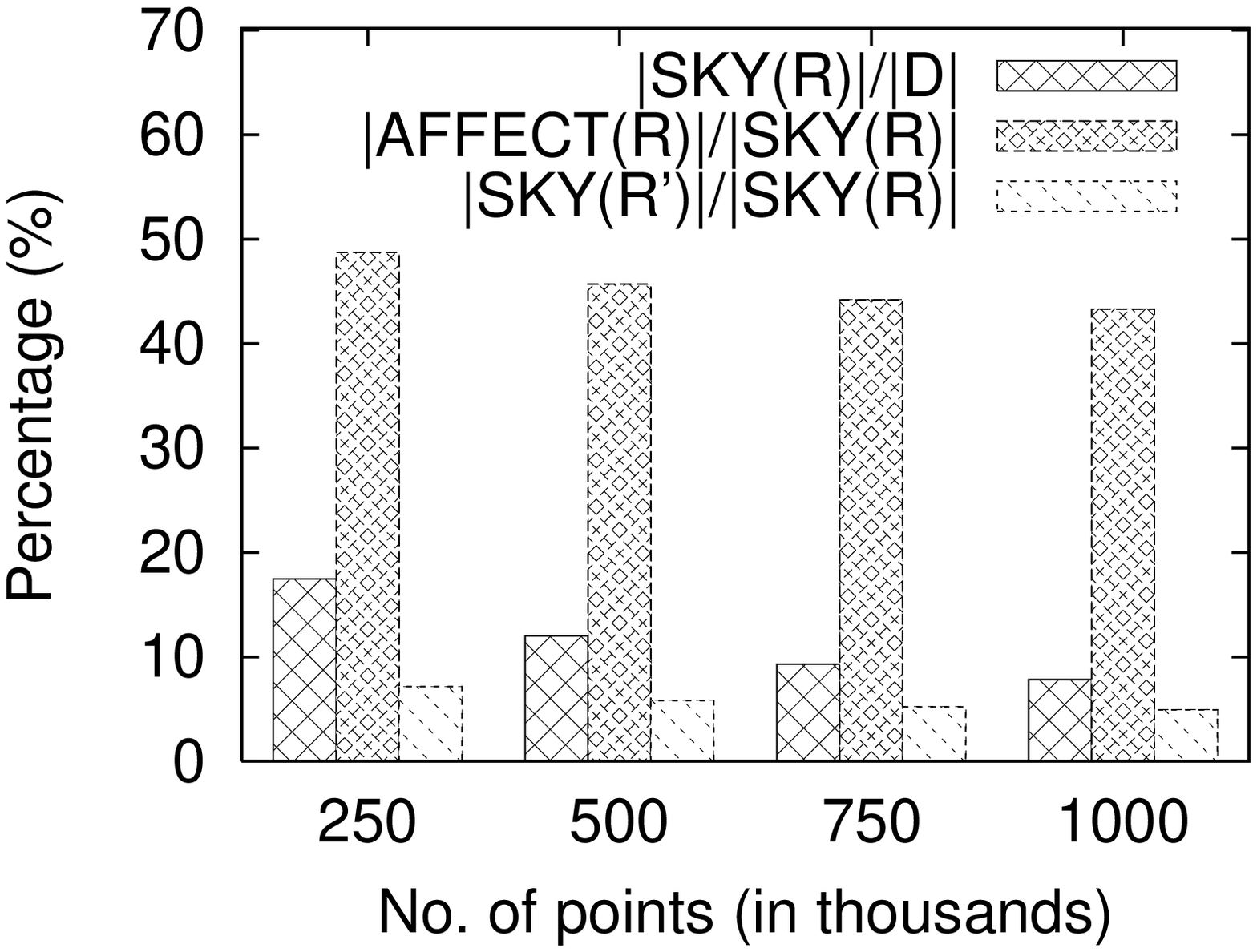}
    \end{minipage}
    \\
(a) &
 (b)
& (c) & (d)
\end{tabular}
\caption{Scalability with respect to database size\vspace*{-0.5cm}}\vspace*{-0.5cm}\label{fig:graphAgainstNoOfTuple}
\end{figure*}

\begin{figure*}[tbp] 
\begin{tabular}{c c c c}
    \begin{minipage}[htbp]{4.0cm}
        \includegraphics[width=4.0cm,height=3.0cm]{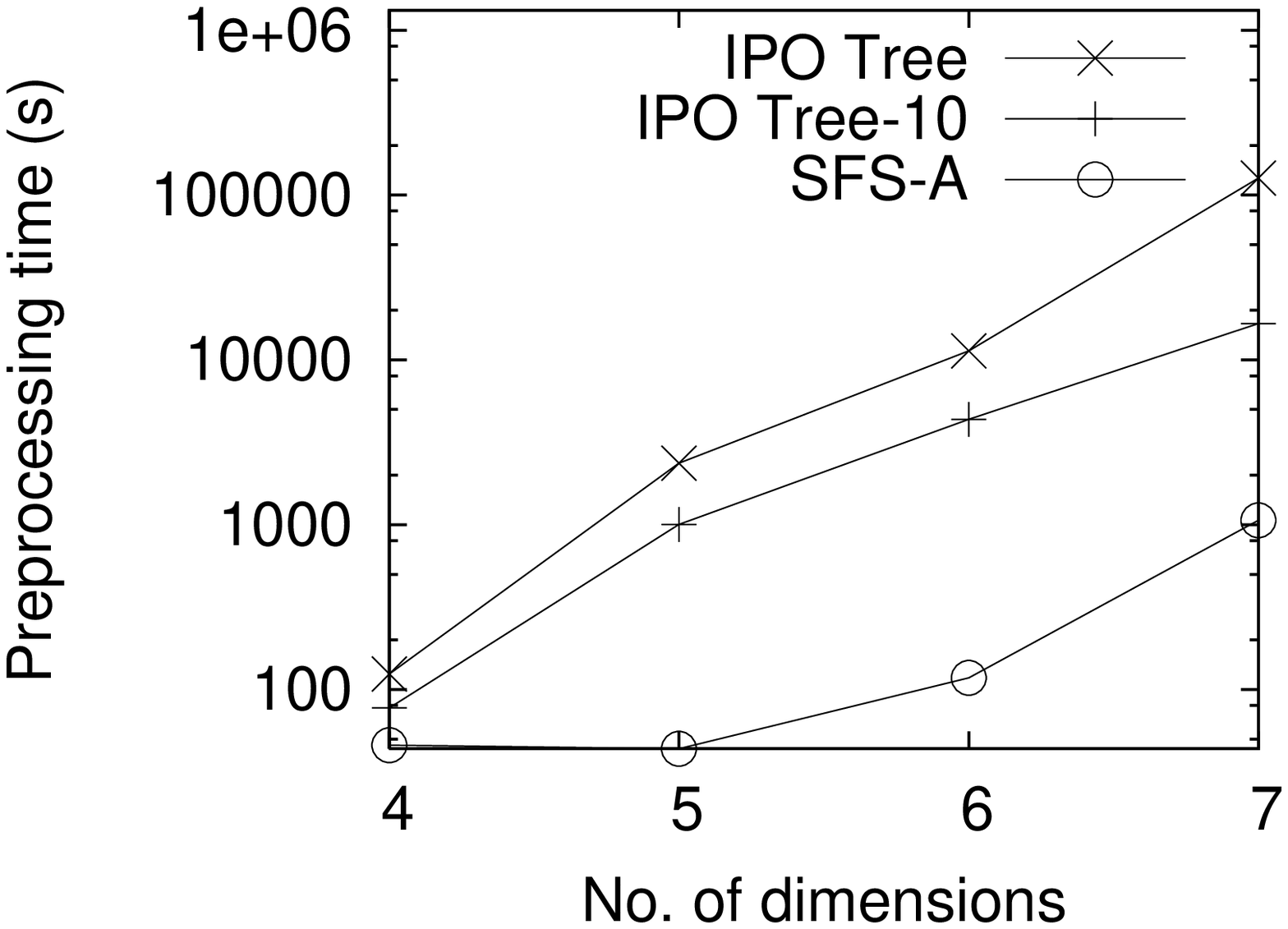}
    \end{minipage}
&
    \begin{minipage}[htbp]{4.0cm}
        \includegraphics[width=4.0cm,height=3.0cm]{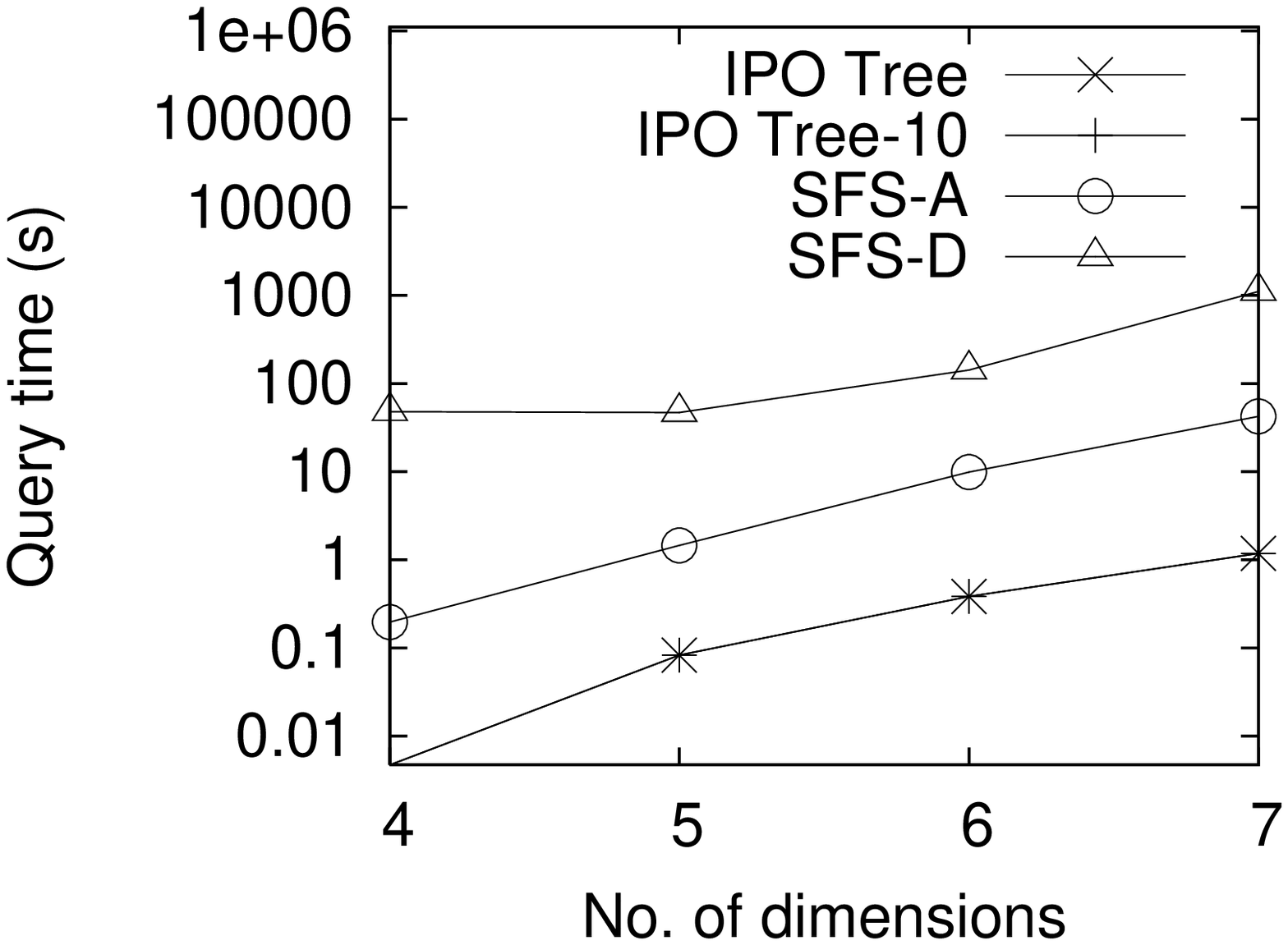}
    \end{minipage}
&
    \begin{minipage}[htbp]{4.0cm}
        \includegraphics[width=4.0cm,height=3.0cm]{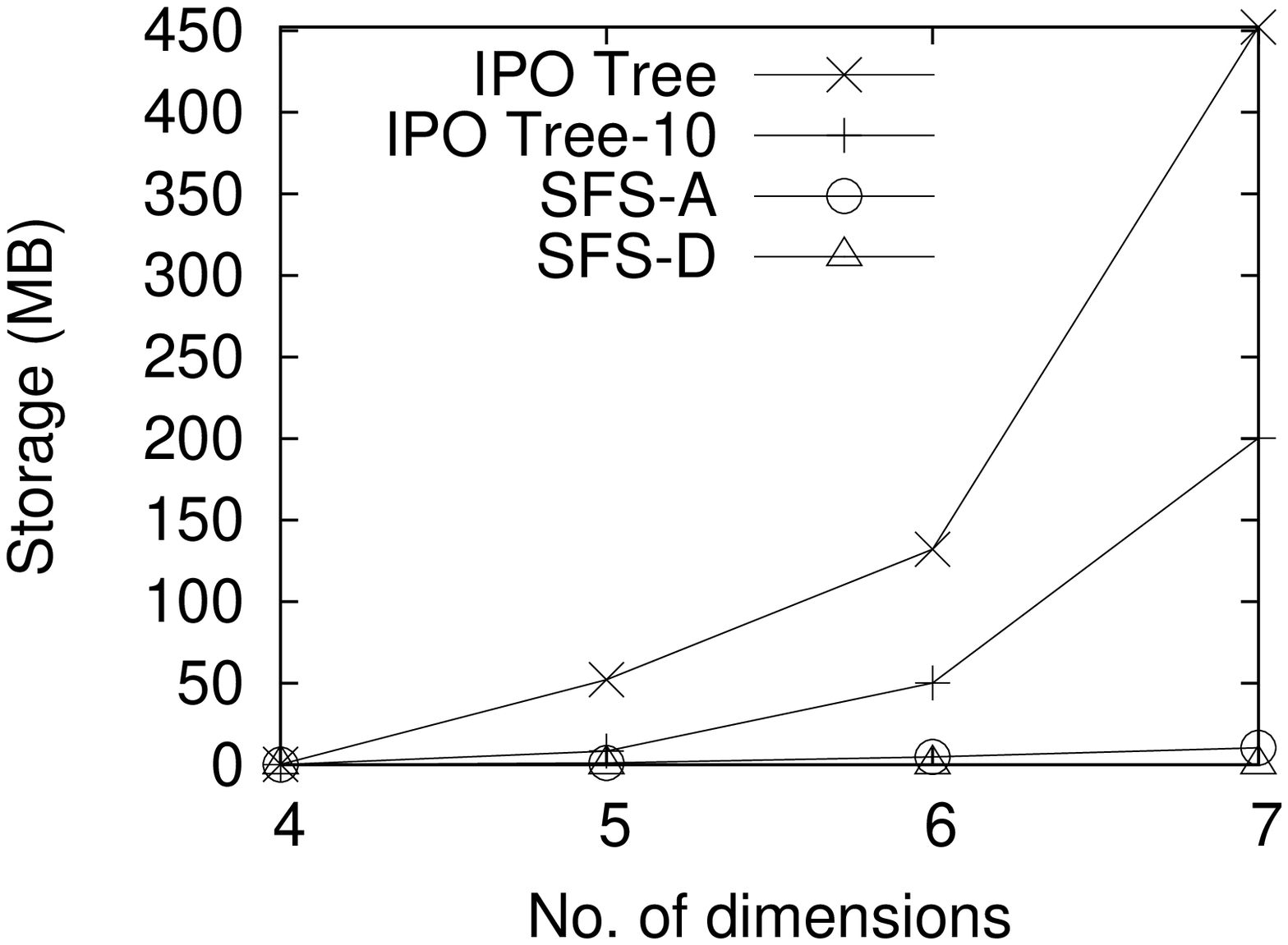}
    \end{minipage}
&
    \begin{minipage}[htbp]{4.0cm}
        \includegraphics[width=4.0cm,height=3.0cm]{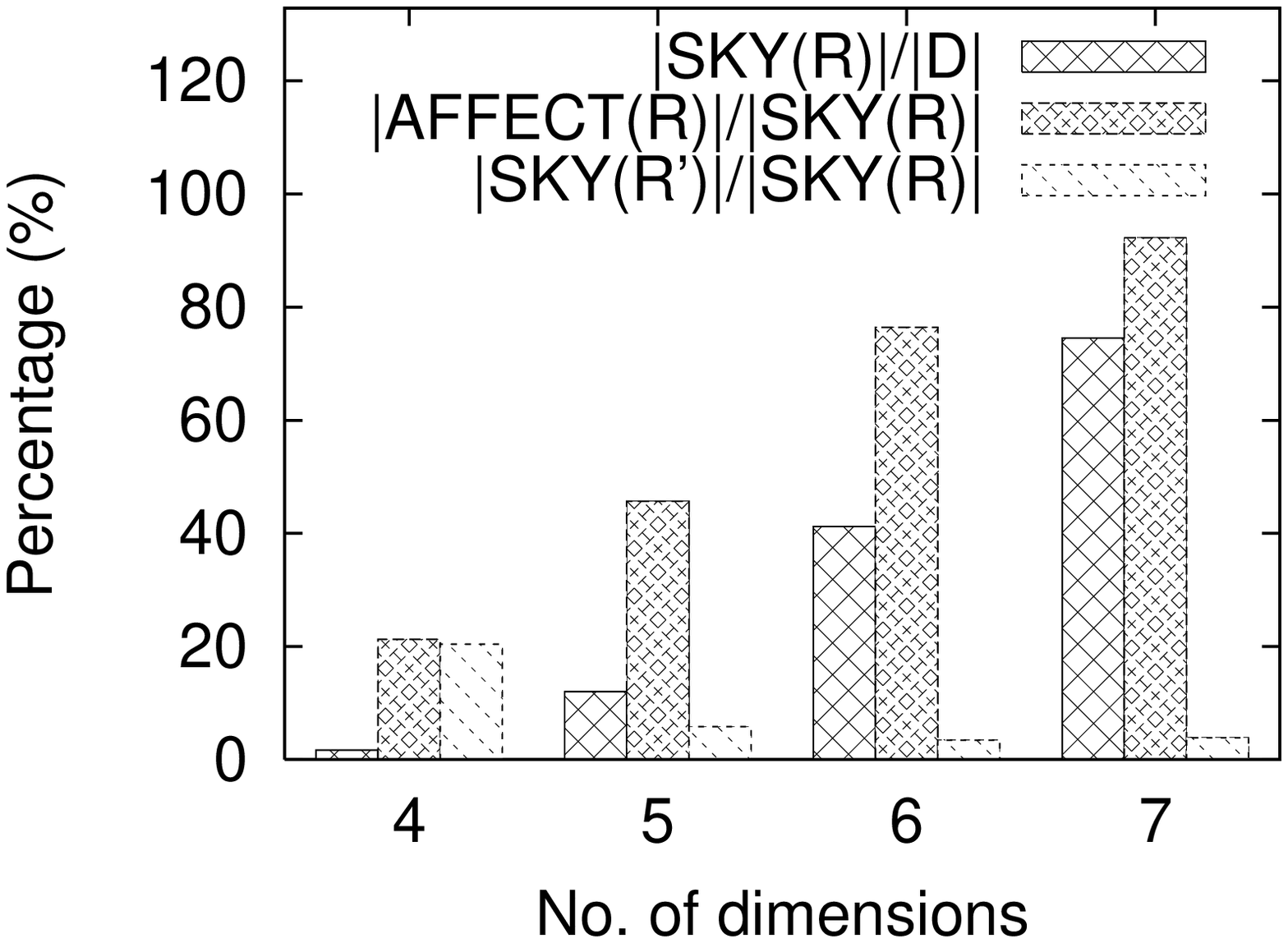}
    \end{minipage}
    \\
(a) &
 (b)
& (c) & (d)
\end{tabular}
\caption{Scalability with respect to dimensionality where no. of numeric attributes is fixed to 3\vspace*{-0.5cm}}\vspace*{-0.5cm}
\label{fig:graphAgainstNoOfCatAttr}
\end{figure*}

\begin{figure*}[tbp] 
\begin{tabular}{c c c c}
    \begin{minipage}[htbp]{4.0cm}
        \includegraphics[width=4.0cm,height=3.0cm]{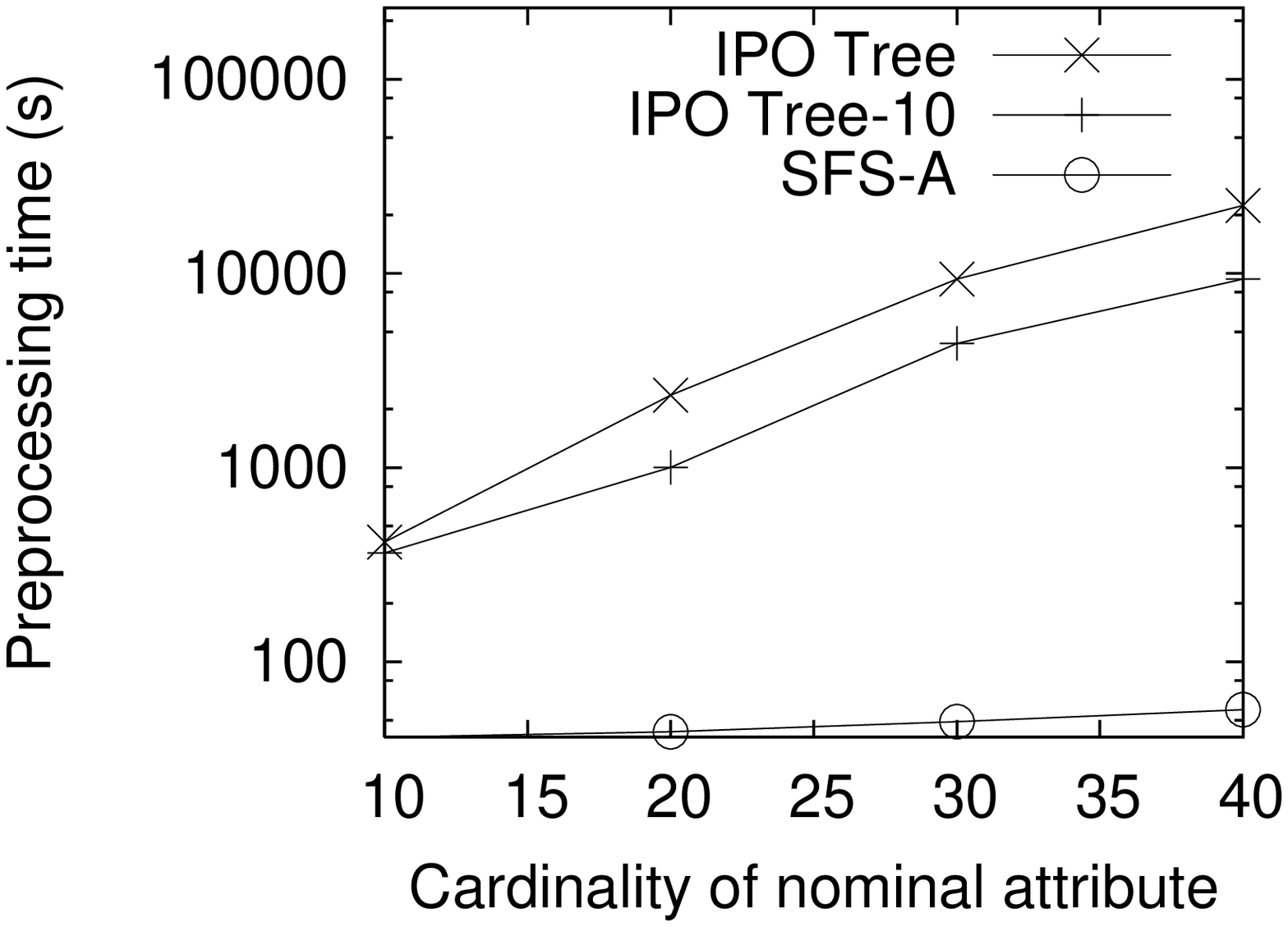}
    \end{minipage}
&
    \begin{minipage}[htbp]{4.0cm}
        \includegraphics[width=4.0cm,height=3.0cm]{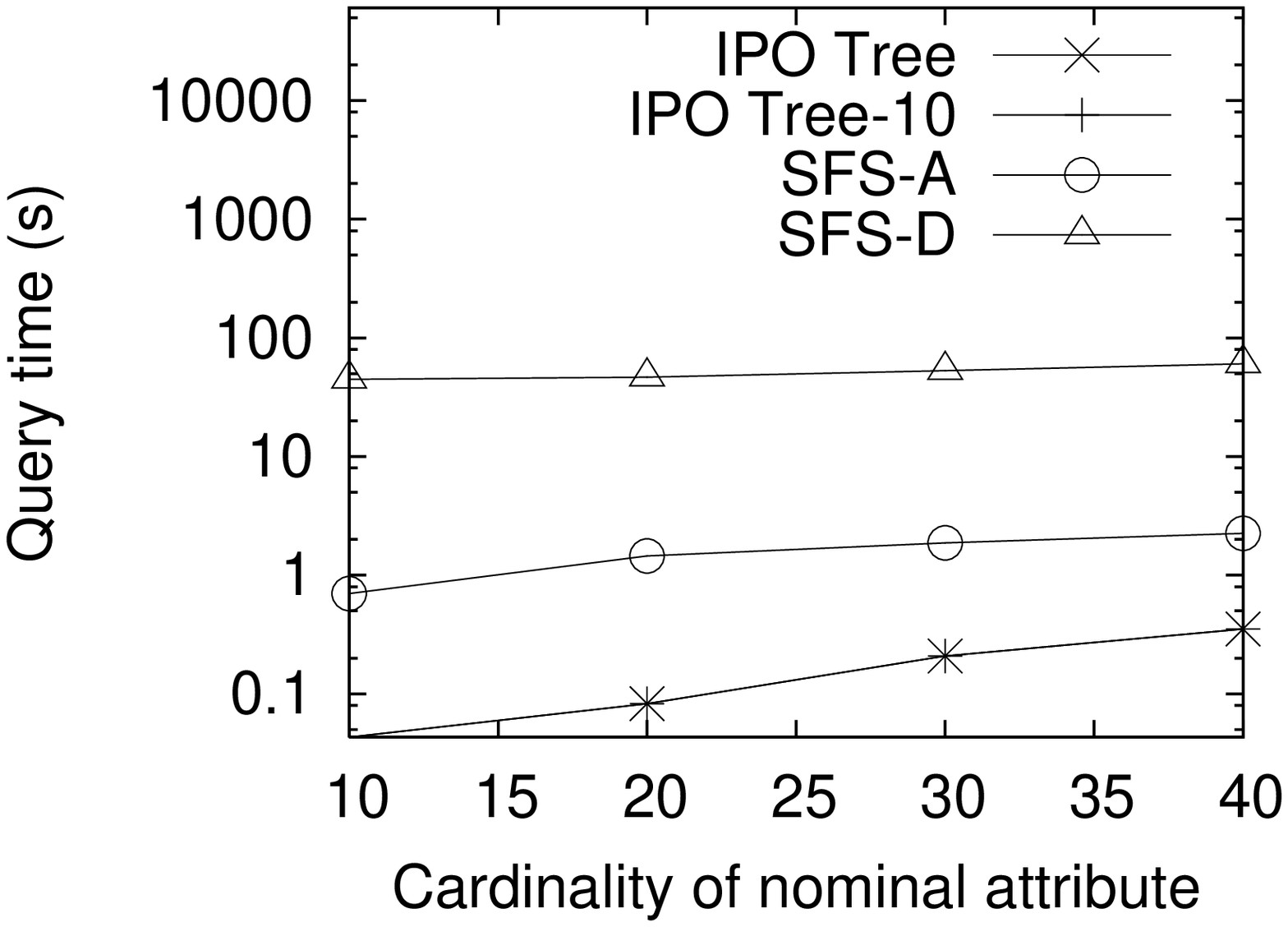}
    \end{minipage}
&
    \begin{minipage}[htbp]{4.0cm}
        \includegraphics[width=4.0cm,height=3.0cm]{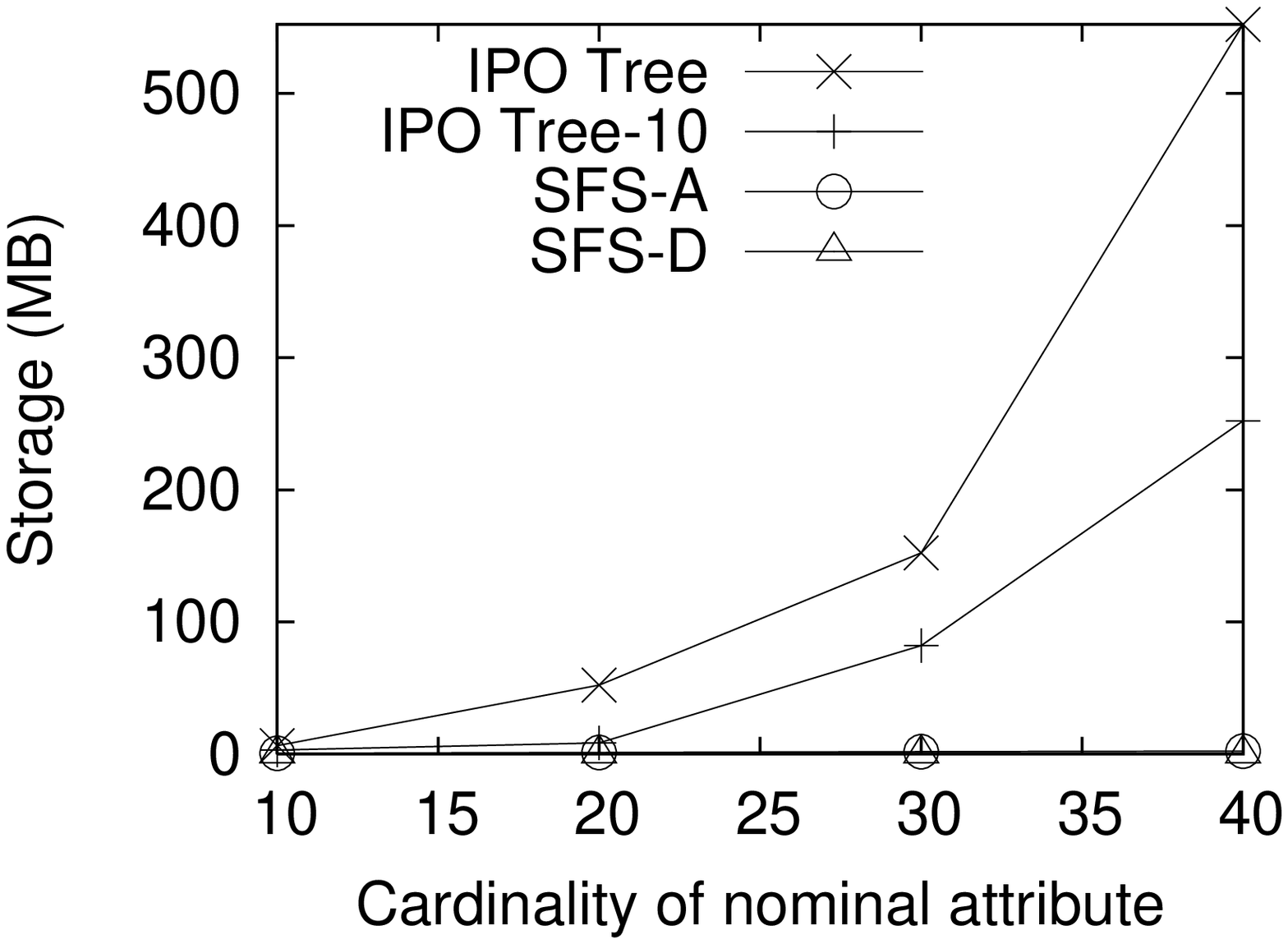}
    \end{minipage}
&
    \begin{minipage}[htbp]{4.0cm}
        \includegraphics[width=4.0cm,height=3.0cm]{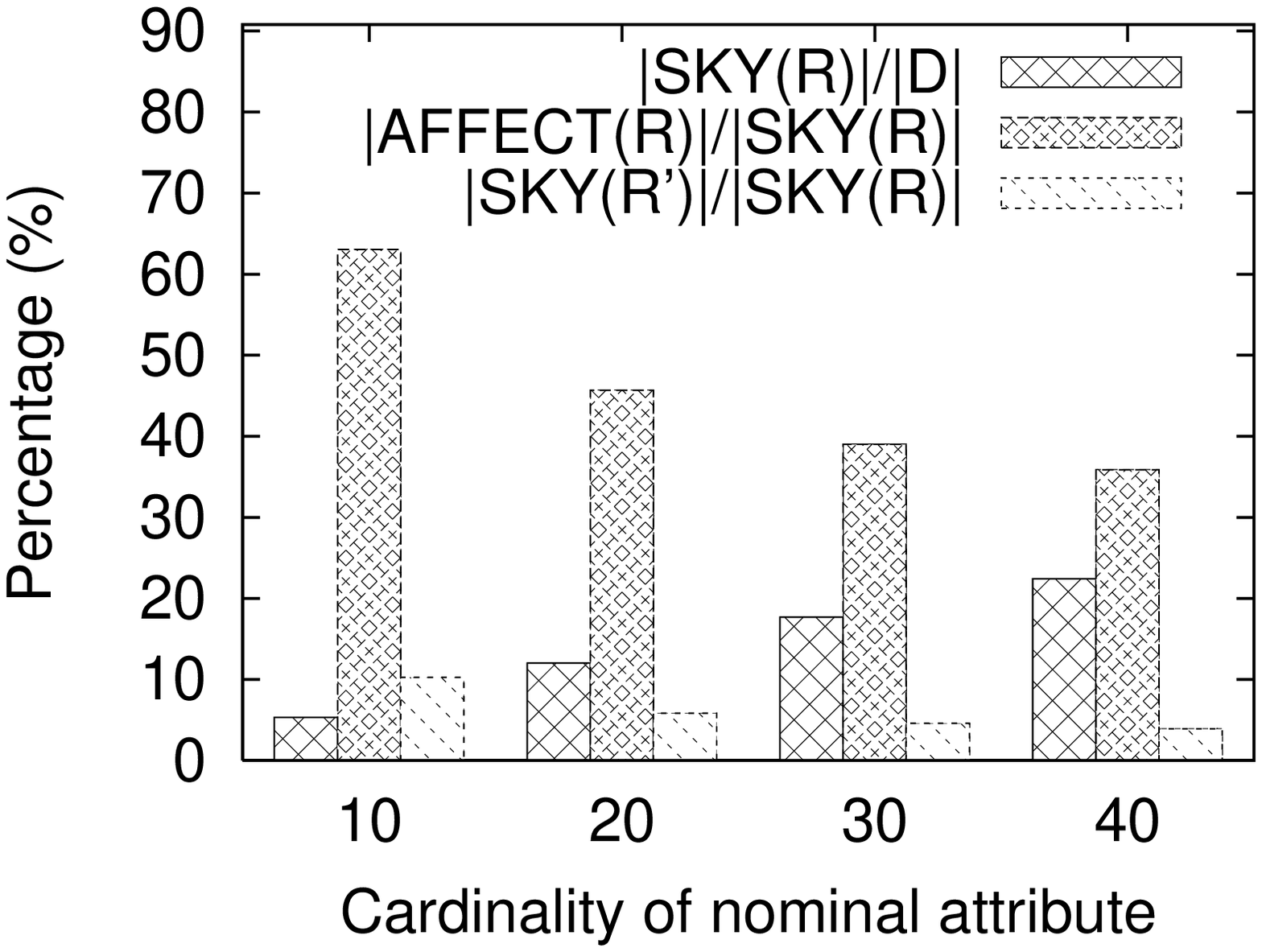}
    \end{minipage}
    \\
(a) &
 (b)
& (c) & (d)
\end{tabular}
\caption{Scalability with respect to cardinality of nominal
attribute\vspace*{-0.5cm}}\vspace*{-0.5cm}\label{fig:graphAgainstCardinality}
\end{figure*}

\begin{figure*}[tbp] 
\begin{tabular}{c c c c}
    \begin{minipage}[htbp]{4.0cm}
        \includegraphics[width=4.0cm,height=3.0cm]{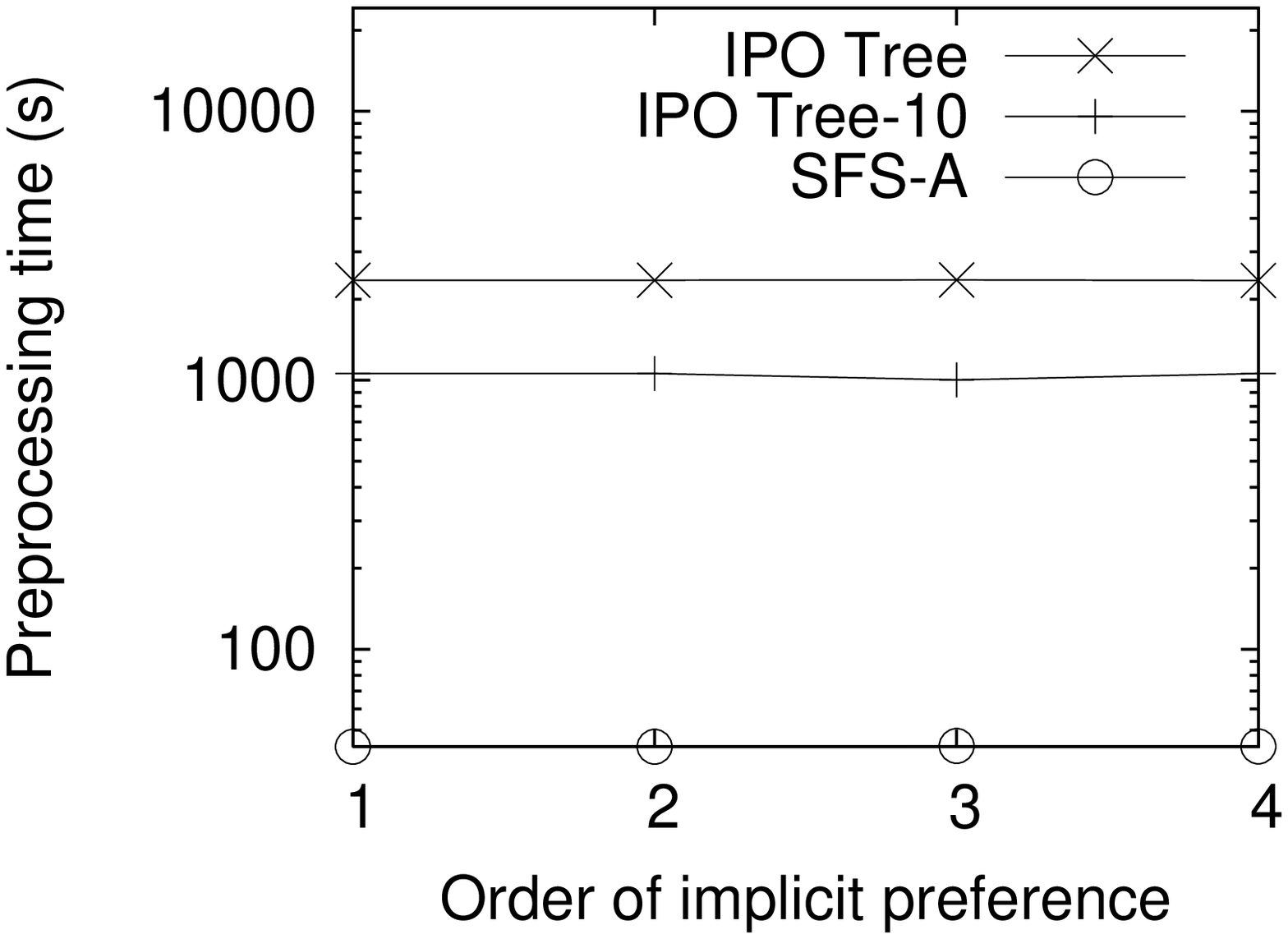}
    \end{minipage}
&
    \begin{minipage}[htbp]{4.0cm}
        \includegraphics[width=4.0cm,height=3.0cm]{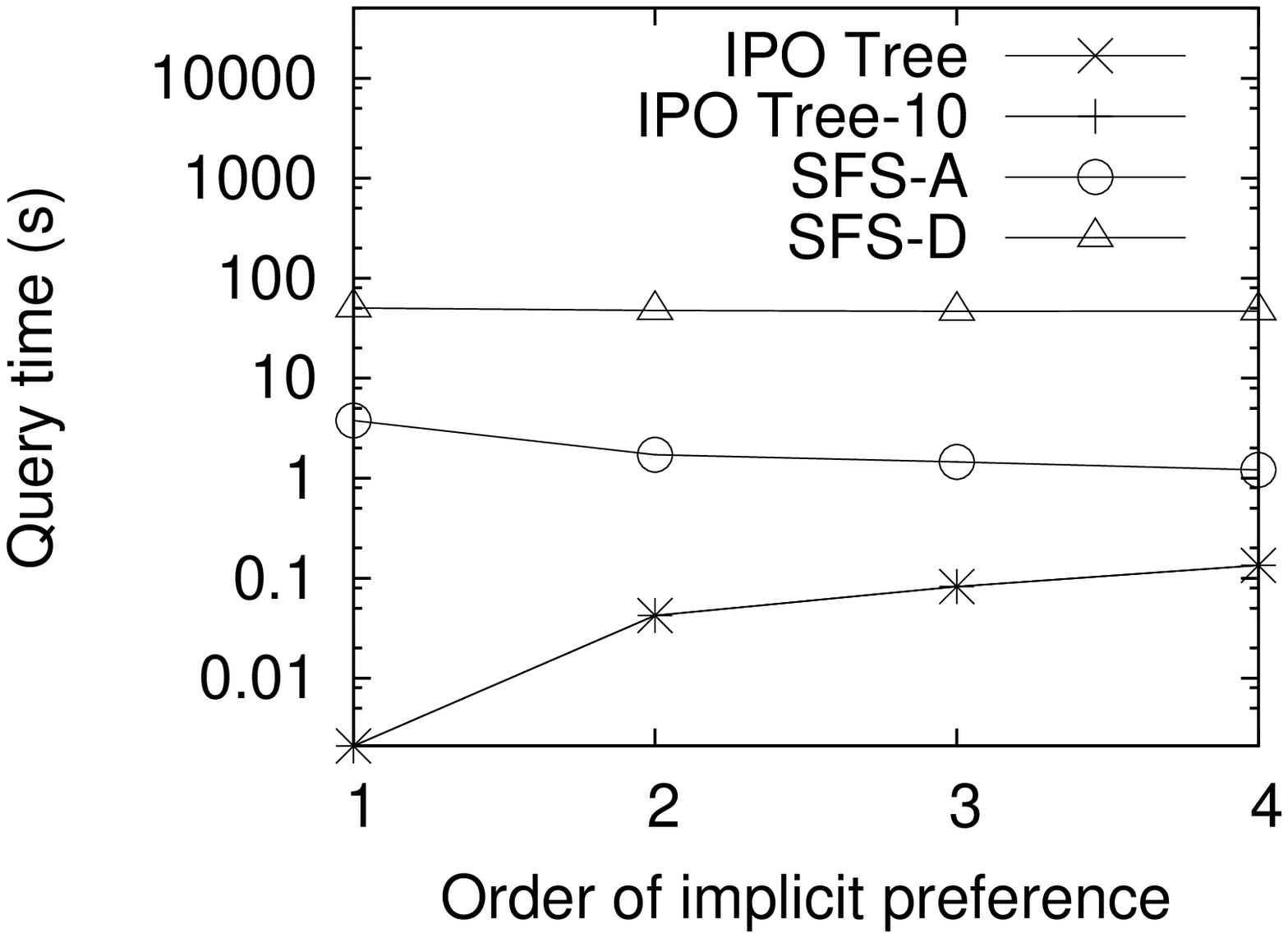}
    \end{minipage}
&
    \begin{minipage}[htbp]{4.0cm}
        \includegraphics[width=4.0cm,height=3.0cm]{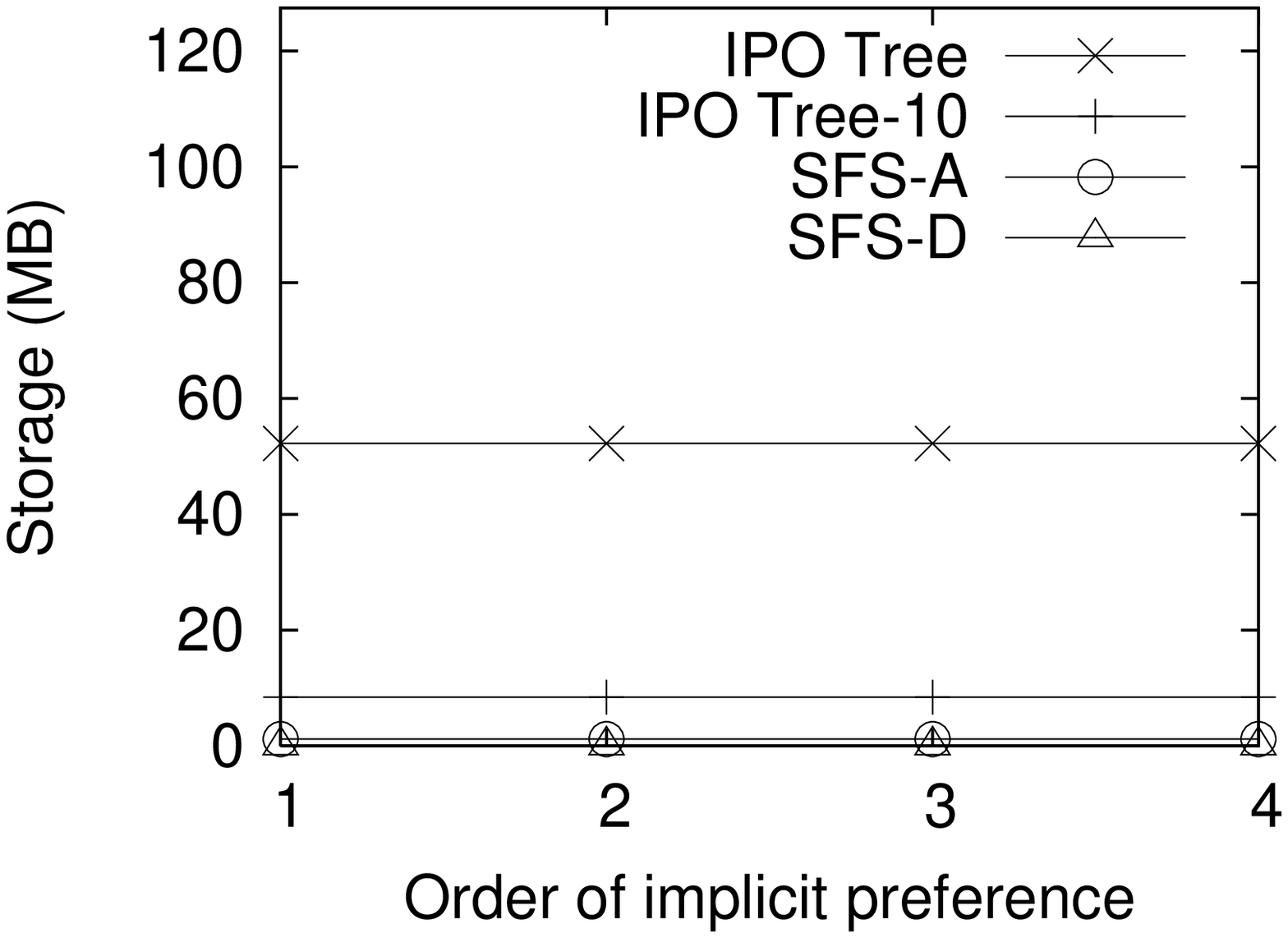}
    \end{minipage}
&
    \begin{minipage}[htbp]{4.0cm}
        \includegraphics[width=4.0cm,height=3.0cm]{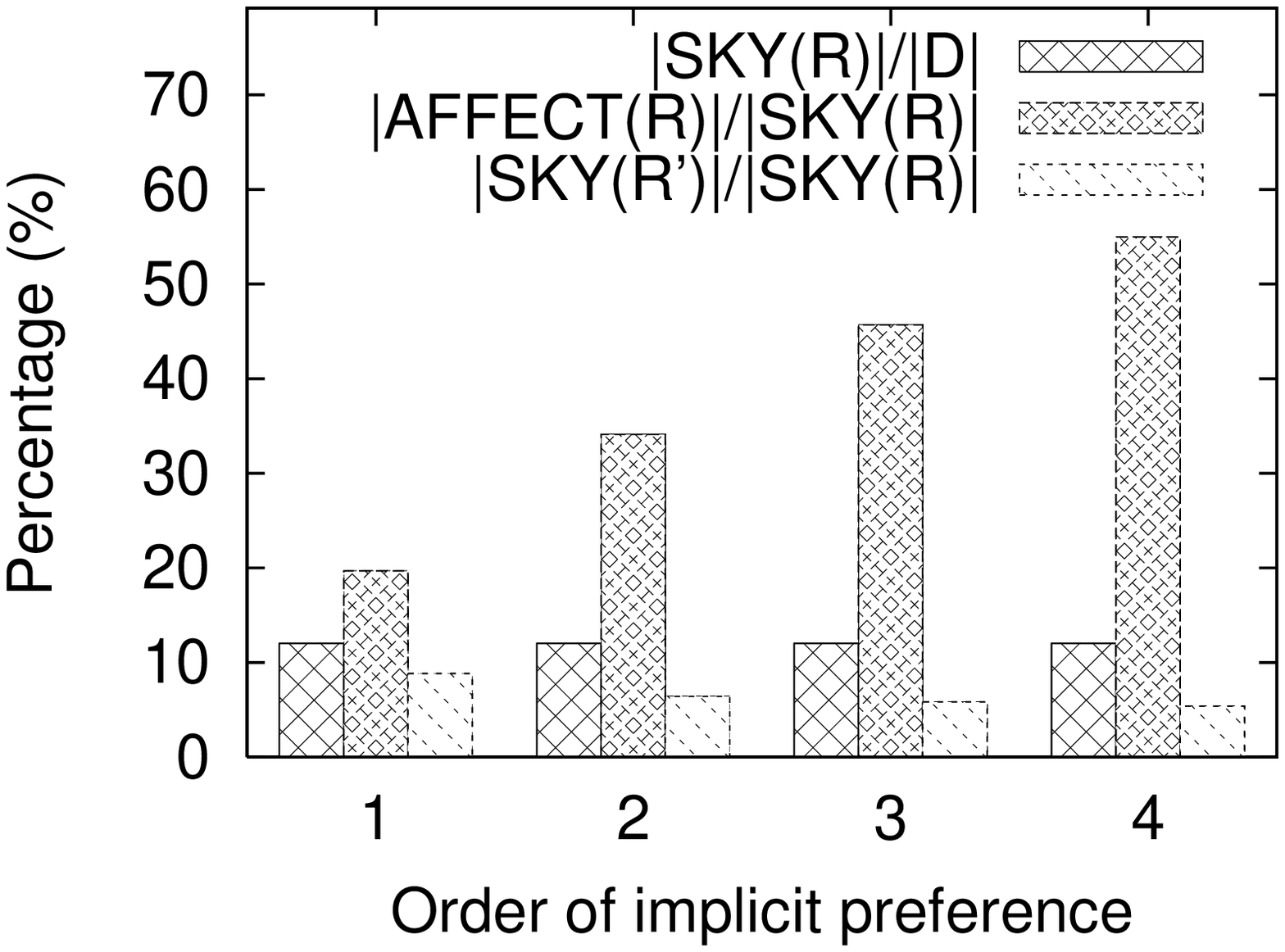}
    \end{minipage}
    \\
(a) &
 (b)
& (c) & (d)
\end{tabular}
\caption{Effect of order of implicit
preference\vspace*{-0.5cm}}\vspace*{-0.5cm}\label{fig:graphAgainstImplicitOrder}
\end{figure*}

\begin{figure*}[tbp] 
\begin{tabular}{c c c c}
    \begin{minipage}[htbp]{4.0cm}
        \includegraphics[width=4.0cm,height=3.0cm]{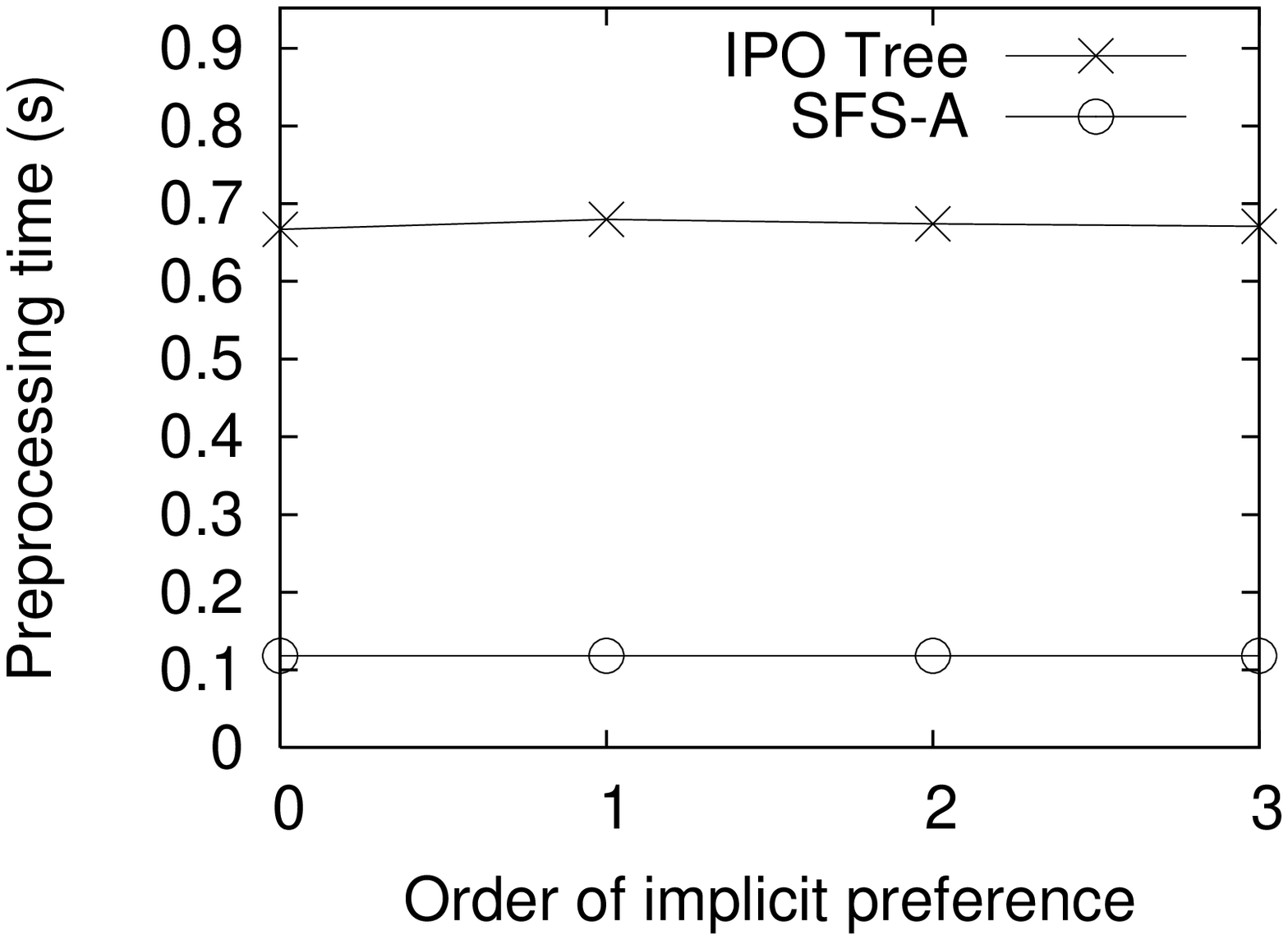}
    \end{minipage}
&
    \begin{minipage}[htbp]{4.0cm}
        \includegraphics[width=4.0cm,height=3.0cm]{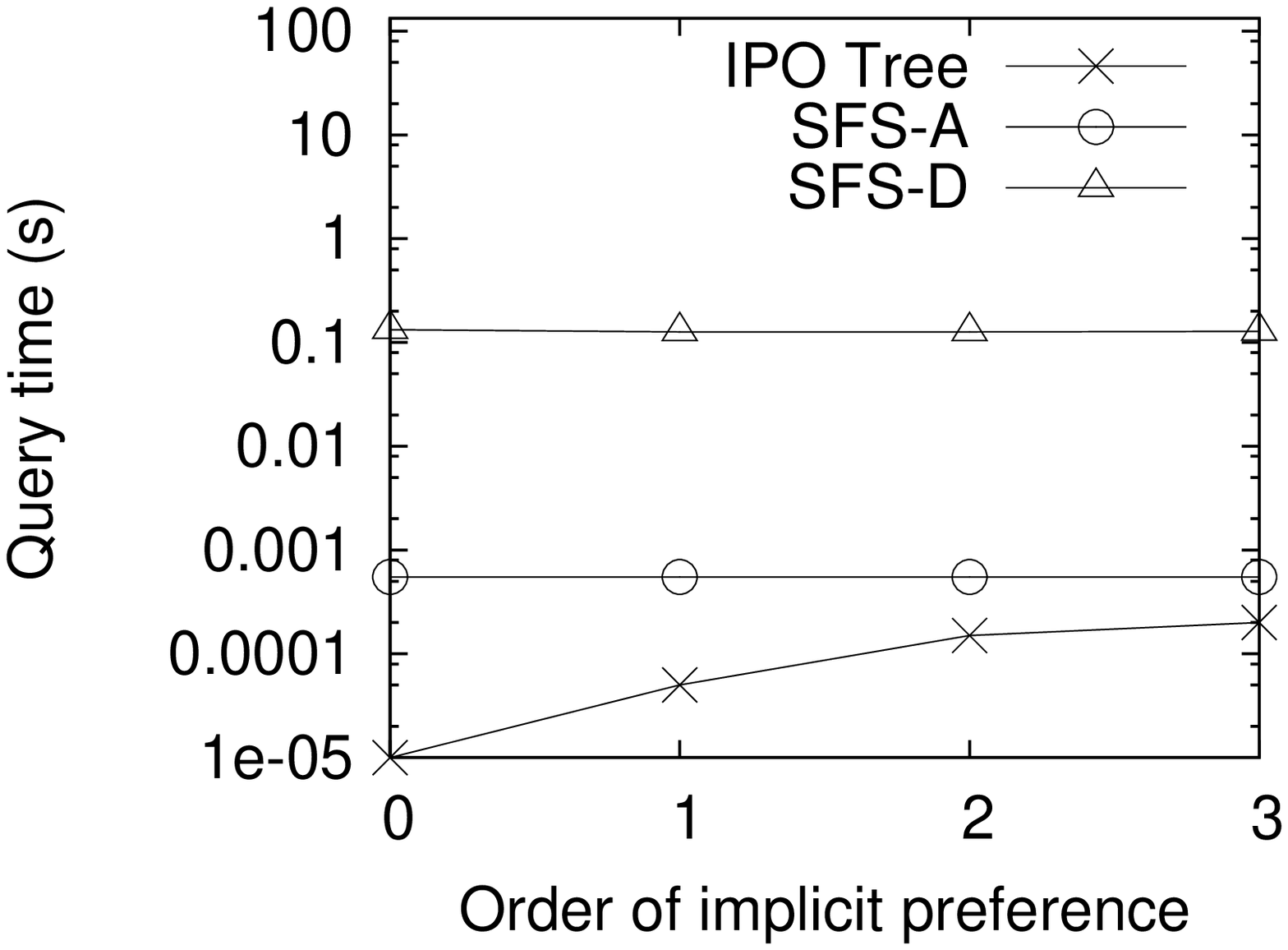}
    \end{minipage}
&
    \begin{minipage}[htbp]{4.0cm}
        \includegraphics[width=4.0cm,height=3.0cm]{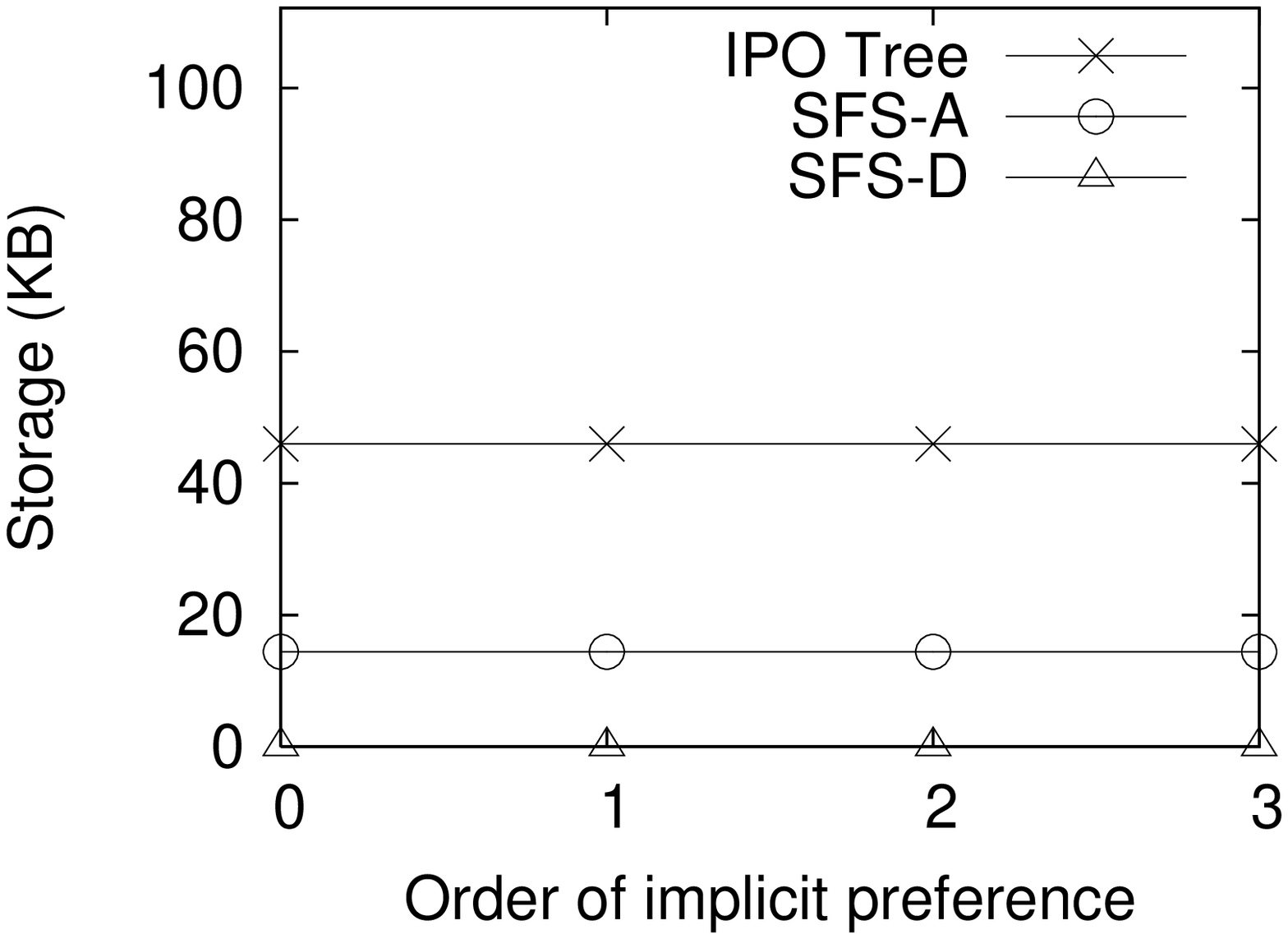}
    \end{minipage}
&
    \begin{minipage}[htbp]{4.0cm}
        \includegraphics[width=4.0cm,height=3.0cm]{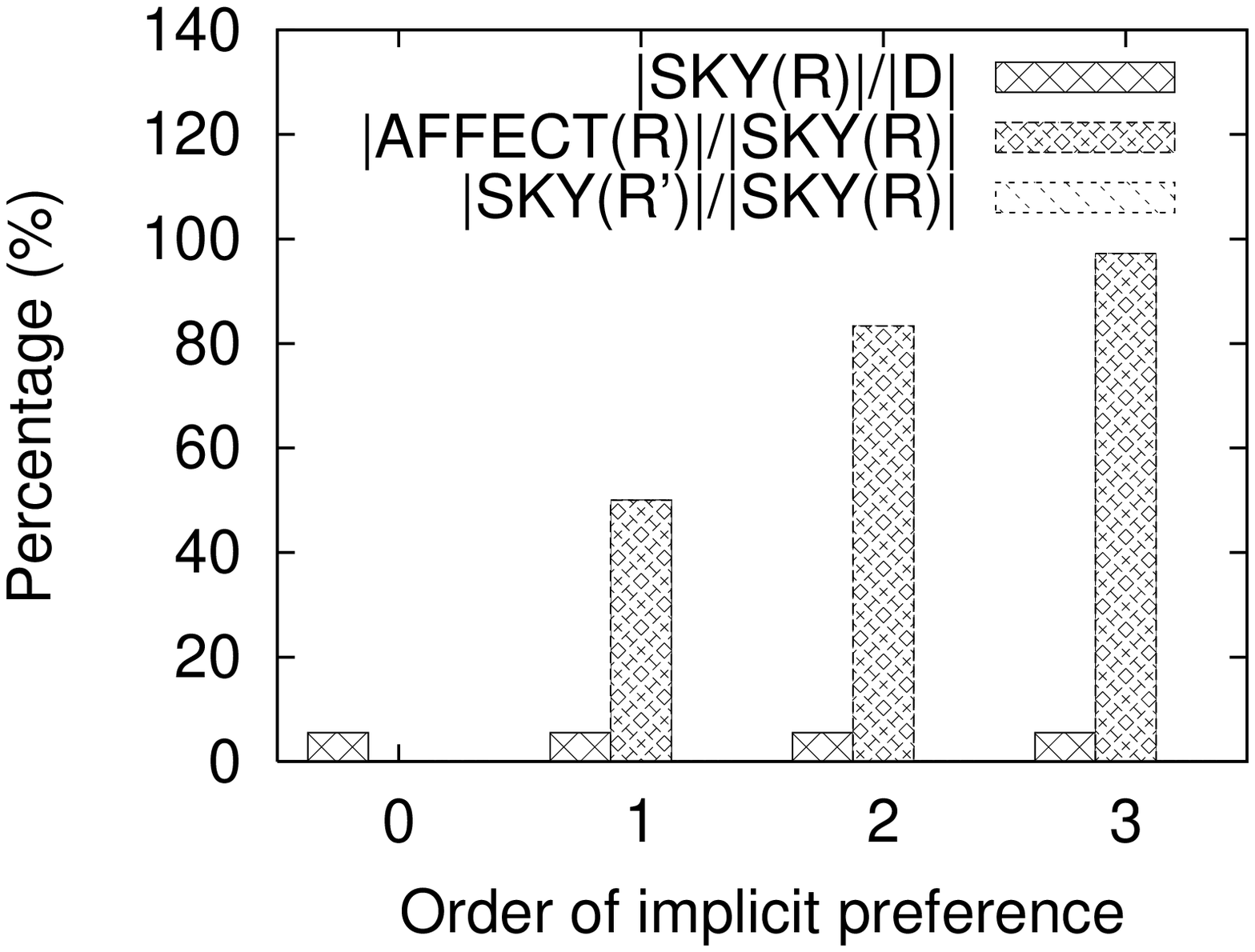}
    \end{minipage}
    \\
(a) &
 (b)
& (c) & (d)
\end{tabular}
\caption{Effect of order of implicit preference (real data
set) 
\vspace*{-0.5cm}}\vspace*{-0.5cm}\label{fig:graphAgainstImplicitOrder-real}
\end{figure*}

\medskip

 \noindent\textit{Effect of the database size:}
In Figure~\ref{fig:graphAgainstNoOfTuple}(d), we note that
$|SKY(R)|/|\mathcal{D}|$ decreases slightly when the data size
increases. This is because, when there are more data points, there is
a higher chance that a data point is dominated by other data points.
Nevertheless, $|SKY(R)|$ increases with database size, and
therefore we see an upward trend in run time and in storage.
For the IPO tree methods, the skyline information size will increase
with data size. For \emph{SFS-A}, the preprocessing time is $O(N log
N + Nn)$ and the query time is $O(l \log n + \min(c, l) \cdot n)$,
where $N$ is the data size, $l = |AFFECT(R)|$, $c = |SKY(R')|$ and
$n = |SKY(R)|$. For \emph{SFS-D} the query time is $O(N log N +
Nn)$. We can see that the results from graphs match with the
complexity expectation.



\medskip

 \noindent\textit{Effect of dimensionality:}
%
%
We study the effect of the number of nominal attribute $m'$ where
the number of numeric attributes is fixed to 3, with the results as shown in
Figure~\ref{fig:graphAgainstNoOfCatAttr}.
 In Figure~\ref{fig:graphAgainstNoOfCatAttr}(d),
$|SKY(R)|/|\mathcal{D}|$ increases. With more nominal attributes, it
is less likely that the data points are dominated by others and thus
$|SKY(R)|$ increases. $|AFFECT(R)|/|SKY(R)|$ also increases with
$m'$ because it is more likely that a data point is affected when
the implicit preference contains preferences on more nominal
attributes. The number of nodes in a full \emph{IPO tree} is given
by $O(c^{m'})$ where $c$ is the cardinality of a nominal attribute.
Because of these factors, the preprocessing time and the query time
of all algorithms increase with $m'$. For the same reason, the
storage for \emph{IPO Tree} and the storage of \emph{SFS-A} also
increase slightly.
%

\medskip

\noindent\textit{Effect of Cardinality of Nominal Attribute:}
Figure~\ref{fig:graphAgainstCardinality}(d) shows that $|SKY(R)|$
increases with cardinality. This is because, when the cardinality
increases, there is a higher chance that a data point is not
dominated by other data points. Also, the number of nodes in a full
\emph{IPO tree} is given by $O(c^{m'})$ where $c$ is the cardinality
of a nominal attribute and $m'$ is the number of nominal attributes.
Thus, the preprocessing time, query time and storage of our proposed
algorithms increases with the cardinality. From
Figure~\ref{fig:graphAgainstCardinality}(b), the increase is
dampened for \emph{SFS-A} because the query time of
\emph{SFS-A} depends on $|AFFECT(R)|$ and there is a decrease in
$|AFFECT(R)|/|SKY(R)|$, which is caused by fewer data points with
frequent nominal values when there are more values in a nominal
attribute.

\medskip

\noindent\textit{Effect of Order of Implicit Preference:} For IPO
tree, the number of set operations is given by $O(x^{m'})$ where $x$
is the order of implicit preference. Hence, in
Figure~\ref{fig:graphAgainstImplicitOrder}(b), the query time for
\emph{IPO Tree} increases. The query times for \emph{SFS-A} and
\emph{SFS-D} are slightly dropping because the skyline size
decreases when the order of implicit preference increases. It is
obvious that neither the pre-processing or storage will be affected.
Figure~\ref{fig:graphAgainstImplicitOrder}(d) shows that the size of
affected skyline points increases. This is because more nominal
values involved in the preference affect more data points.

\subsection{Real Data Set}

To demonstrate the usefulness of our methods,
we ran our algorithms on a real data set,
Nursery data set, which is publicly available from
the UCIrvine Machine Learning Repository\footnote{{\tt
http://kdd.ics.uci.edu/}}.
In this data set, there are 12,960 instances and 8
attributes.
The experimental setup is same as \cite{WPF+07a}.
There are six totally-order attributes and two nominal attributes,
namely form of the family and the number of
children. (Note that although the number of
children is a numeric attribute, it is not clear whether a family
with one child is ``better" than a family with two children.)
The
cardinality of both nominal attributes are equal to 4.
%
The results in the performance are similar to those for
the synthetic data sets.
Figure~\ref{fig:graphAgainstImplicitOrder-real} shows the results on the
real data set with the effect of the order of implicit preference.


\medskip

\subsection{Main Observations}
The major findings from the experiments are the followings. The
\emph{SFS-D} algorithm cannot meet real-time requirements, since the
query time is at least in terms of tens of seconds and, in some
cases, exceeds 1000 seconds. In general, \emph{IPO Tree} is the
fastest but \emph{SFS-A} can also return the result within a second
in most cases and under 20 seconds in the worst case, and is orders
of magnitude faster than \emph{SFS-D}. The results with \emph{IPO
Tree-10} show that, by handling a smaller set of nominal values, one
can control both the pre-processing and storage costs. A hybrid
approach adopting \emph{IPO Tree} for popular values and
\emph{SFS-A} for handling queries involving the remaining values is
a sound solution.

\section{Conclusion}
Most previous works on the skyline problem consider data sets with
attributes following a fixed ordering. However, nominal attributes
with dynamic orderings according to different users exist in almost
all conceivable real-life applications. In this work, we study the
problem of online response for such dynamic preferences, two methods
are proposed with different flavors: a semi-materialization method
and an adaptive SFS method.
Our experiments show how our proposed algorithms are
useful in different problem settings.

\bibliographystyle{latex8}
\small
\bibliography{freesky}

\section{Appendix: Proof of Theorem~\ref{thm:merging2}}

%
%

\noindent\textbf{Proof:}
%
%
%
We need to show that a point $p$ is in $SKY(\widetilde{R}''')$ if
and only if it is in $(SKY(\widetilde{R}') \cap
SKY(\widetilde{R}'')) \cup PSKY(\widetilde{R}')$. For each
direction, we prove by contradiction.

[A] Firstly, assume $p$ is in $SKY(\widetilde{R}''')$, 
and suppose that $p$ is not in $(SKY(\widetilde{R}') \cap
SKY(\widetilde{R}'')) \cup PSKY(\widetilde{R}')$. Then, 
%
%
by Theorem~\ref{thm:mono}, since $p \in SKY(\widetilde{R}''')$ and
$\widetilde{R}'''$ is a refinement of $\widetilde{R}'$, we deduce
that $p \in SKY(\widetilde{R}')$. Thus, 
$p$ must satisfy the following:

$\bullet$ Condition 1: $p.D_i \not\in \{v_1, ... v_{x-1}\}$ and

$\bullet$ Condition 2: $p \not\in SKY(\widetilde{R}'')$.

Consider Condition 2. Since $p \not\in SKY(\widetilde{R}'')$, there
exists a data point $q$ dominating $p$ w.r.t $\widetilde{R}''$.
In other words, with respect to $\widetilde{R}''$, $q.D_k \preceq p.D_k$ for
all $k$ and in at least one dimension $D_j$,  $q.D_j \prec
p.D_j$. Let $\mathcal{J}$
be the set of dimensions $D_j$ where $q.D_j \prec p.D_j$ w.r.t $\widetilde{R}''$. Besides,
for all dimensions $D_k$ other than $D_i$, the partial
orders of $\widetilde{R}''$ and $\widetilde{R}'''$ are the same.
Hence, w.r.t.
$\widetilde{R}'''$, $q.D_k \preceq p.D_k$ for all $k (\neq i)$.
There are two subcases: \emph{Case (i)}: $D_i \not\in \mathcal{J}$ and 
\emph{Case (ii)}: $D_i \in \mathcal{J}$. 

\emph{Case (i)}: $D_i \not\in \mathcal{J}$.
For all $D_j \in \mathcal{J}$, since $q.D_j \prec p.D_j$ w.r.t $\widetilde{R}''$
and the partial orders in $\widetilde{R}''_j$
are those in $\widetilde{R}'''_j$, we have
$q.D_j \prec p.D_j$ w.r.t.
$\widetilde{R}'''$.
Also, w.r.t. $\widetilde{R}'''$, $q.D_k \preceq p.D_k$ for all $k
\neq i$. Hence, since $i \not\in \mathcal{J}$, for dimension $D_i$,
it must be the case that $p.D_i \prec q.D_i$ w.r.t
$\widetilde{R}'''$. Otherwise, $p$ is dominated by $q$ w.r.t
$\widetilde{R}'''$, and $p$ cannot be in $SKY(\widetilde{R}''')$.
Since $p.D_i \prec q.D_i$ w.r.t $\widetilde{R}'''$, we have $p.D_i
\neq q.D_i$. Since $q.D_k \preceq p.D_k$ w.r.t. $\widetilde{R}''$
for all $k$, and $p.D_i \neq q.D_i$, we have $q.D_i \prec p.D_i$
w.r.t $\widetilde{R}''$. Since the implicit preference in
$\widetilde{R}''$ is ``$v_x \prec *$", we conclude that $p.D_i$
cannot be $v_x$. Since $\widetilde{R}'''$ is ``$v_1 \prec ... \prec
v_x \prec *$" and $p.D_i \prec q.D_i$ w.r.t $\widetilde{R}'''$,
$p.D_i$ must be in $\{v_1, ... v_{x-1}\}$.
However, this violates Condition 1 discussed above. 
Hence, we arrive at a
contradiction.

\emph{Case (ii)}: $D_i \in \mathcal{J}$. 
We obtain $q.D_i \prec p.D_i$ w.r.t.
$\widetilde{R}''$.
Besides, since the implicit preference in $\widetilde{R}''$
is ``$v_x \prec *$", $q.D_i$ must be equal to $v_x$ and $p.D_i$ cannot
be equal to $v_x$. Since $p \in SKY(\widetilde{R}''')$, there is no
other point including $q$ dominating $p$ w.r.t. $\widetilde{R}'''$.
Note that, w.r.t.
$\widetilde{R}'''$, $q.D_k \preceq p.D_k$ for all $k (\neq i)$.
We obtain $p.D_i \preceq q.D_i$ w.r.t. $\widetilde{R}'''$.
(Otherwise, $q.D_i \prec p.D_i$ w.r.t $\widetilde{R}'''$
and $p$ is dominated by $q$ w.r.t. $\widetilde{R}'''$, which
leads to a contradiction.)
Besides, since  $q.D_i
= v_x$, $p.D_i \neq v_x$ and $\widetilde{R}'''$ is ``$v_1 \prec ... \prec v_x \prec *$",
$p.D_i$ must be in $\{v_1, ... v_{x-1}\}$. However, this
violates
Condition 1. 
Hence, we arrive at a contradiction.

[B] Conversely, 
consider a point $p$ in $(SKY(\widetilde{R}') \cap
SKY(\widetilde{R}'')) \cup PSKY(\widetilde{R}')$. Suppose that $p$
is not in $SKY(\widetilde{R}''')$. Thus, $p$ is dominated by some
point $q$ w.r.t. $\widetilde{R}'''$. That is, w.r.t
$\widetilde{R}'''$, $q.D_k \preceq p.D_k$ for all $k$ and $q.D_j
\prec p.D_j$ for at least one dimension $D_j$.

Since $p \in (SKY(\widetilde{R}') \cap SKY(\widetilde{R}'')) \cup
PSKY(\widetilde{R}')$, we know that at least one of the
following two conditions holds.

$\bullet$ Condition 3: $p.D_i \in \{v_1,...v_{x-1}\}$ and $p \in
SKY(\widetilde{R}')$, or

$\bullet$ Condition 4: 
$p \in SKY(\widetilde{R}')$ and $p \in SKY(\widetilde{R}'')$.

Consider Condition 3.
 Since $p \in SKY(\widetilde{R}')$ and $p \not\in SKY(\widetilde{R}''')$
 where $\widetilde{R}'''_i$ is a refinement of $\widetilde{R}'_i$, and
 $\widetilde{R}'''_k = \widetilde{R}'_k$ for all $k \neq i$,
we deduce that $q.D_i \prec p.D_i$ exists in partial orders of $\widetilde{R}'''$
but not in partial orders of $\widetilde{R}'$.
Since $q.D_i \prec p.D_i$ 
w.r.t.
$\widetilde{R}'''$, $p.D_i \in \{v_1, ..., v_{x-1}\}$
and $\widetilde{R}'''$ is ``$v_1 \prec ... \prec v_x \prec *$", we deduce
$q.D_i \in \{v_1, ...v_{x-2}\}$. 
For each possible binary order $q.D_i \prec p.D_i$ w.r.t. $\widetilde{R}'''$ where
$p.D_i \in \{v_1, ..., v_{x-1}\}$ and $q.D_i \in \{v_1, ...v_{x-2}\}$,
we also conclude that $q.D_i \prec p.D_i$ exists in the partial
orders of $\widetilde{R}'$, which leads to a contradiction.

Consider Condition 4.
Since $\widetilde{R}'$, $\widetilde{R}''$ and
$\widetilde{R}'''$ differ only at dimension $D_i$, we only need to
check their implicit preferences to see that, whenever $q.D_i \preceq
p.D_i$ (or $q.D_i \prec p.D_i$) w.r.t. $\widetilde{R}'''$,
it is also true w.r.t. $\widetilde{R}'$ or
$\widetilde{R}''$. Therefore, $q$ also dominates $p$ w.r.t.
$\widetilde{R}'$ or $\widetilde{R}''$.
That is, $p \not\in SKY(\widetilde{R}')$ or $p \not\in SKY(\widetilde{R}'')$, which
leads to a contradiction.
 \done

\end{document}